%% file: paper_dynamic_only.tex
\documentclass[twoside,leqno,twocolumn]{article}

\usepackage{ltexpprt}
\usepackage{booktabs}
\usepackage{hyperref}
\usepackage{tikz}
\usepackage{pdflscape}
\usepackage{microtype}

\usepackage{amsfonts} 
\usepackage{mathtools}
\usepackage{graphicx}
\usepackage{pdfpages}
\usepackage{algorithm}
\usepackage[noend]{algpseudocode}
\usepackage{xspace}

\usepackage{booktabs}
\usepackage[autolanguage]{numprint}
\newcommand{\Id}[1]{\texttt{\detokenize{#1}}}

\usepackage{todonotes}

\input{shortings}

\input{math_definitions.tex}

\input{makros.tex}

\renewcommand{\term}[1]{\textit{#1}}
\author{Jannick Borowitz \thanks{Heidelberg University.}
\and Ernestine Großmann \thanks{Heidelberg University.}
\and Christian Schulz \thanks{Heidelberg University.}}

\begin{document}

\newcommand\relatedversion{}

\title{Optimal Neighborhood Exploration for Dynamic Independent Sets}

\date{}

\maketitle

\fancyfoot[R]{\scriptsize{Copyright \textcopyright\ 20XX by SIAM\\
Unauthorized reproduction of this article is prohibited}}

\begin{abstract} %
A dynamic graph algorithm is a data structure that supports edge insertions, deletions, and specific problem queries. While extensive research exists on dynamic algorithms for graph problems solvable in polynomial time, most of these algorithms have not been implemented or empirically evaluated.

This work addresses the NP-complete maximum weight and cardinality independent set problems in a dynamic setting, applicable to areas like dynamic map-labeling and vehicle routing. Real-world instances can be vast, with millions of vertices and edges, making it challenging to find near-optimal solutions quickly. Exact solvers can find optimal solutions but have exponential worst-case runtimes. Conversely, heuristic algorithms use local search techniques to improve solutions by optimizing vertices. 

In this work, we introduce a novel local search technique called \textit{optimal neighborhood exploration}. This technique creates independent subproblems that are solved to optimality, leading to improved overall solutions. Through numerous experiments, we assess the effectiveness of our approach and compare it with other state-of-the-art dynamic solvers. Our algorithm features a parameter, the subproblem size, that balances running time and solution quality. With this parameter, our configuration matches state-of-the-art performance for the cardinality independent set problem.
By increasing the parameter, we significantly enhance solution quality. 

\end{abstract}

\section{Introduction}

\setcounter{page}{1}%
\pagenumbering{arabic}

Complex graphs are useful in a wide range of applications from technological networks to biological systems like the human brain.
These graphs can contain billions of vertices and edges. 
In practice, the underlying graphs often change over time, \ie, vertices
or edges are inserted or deleted as time is passing.
In a social network, for example, users sign up or leave, and relations between
them may be created or removed over time, or in road networks new roads are built.
Terminology-wise, a problem is said to be \emph{fully dynamic} if the update
operations include both insertions \emph{and} deletions of edges, and
\emph{partially dynamic} if only one type of update operation is allowed.
In this context, a problem is called \emph{incremental}, if only edge
insertions occur, but no deletions, and \emph{decremental}~vice~versa.

A (fully) dynamic graph algorithm is a data structure that supports edge insertions and edge deletions. It also answers certain queries that are specific to the problem under consideration.
The most studied dynamic problems are graph problems such as connectivity, reachability, shortest paths, or matching (see~\cite{DBLP:journals/corr/abs-2102-11169}).  
However, while there is a large body of theoretical work on efficient dynamic graph algorithms, only recently experimental work in the area has been gaining momentum. 
For some classical dynamic algorithms, experimental studies have been performed, such as early works on (all pairs) shortest paths \cite{DBLP:journals/jea/FrigioniINP98,DBLP:journals/talg/DemetrescuI06} or transitive closure~\cite{DBLP:journals/jea/KrommidasZ08} and later contributions for fully dynamic graph clustering~\cite{DBLP:conf/wads/DollHW11}, fully dynamic approximation of betweenness centrality~\cite{DBLP:conf/esa/BergaminiM15} as well as fully dynamic (weighted) matching \cite{DBLP:conf/acda/AngrimanM0U21,DBLP:conf/esa/Henzinger0P020}, fully dynamic delta orientations~\cite{DBLP:conf/acda/BorowitzG023} and fully dynamic minimum cuts \cite{DBLP:conf/alenex/HenzingerN022}. 
However, for other fundamental dynamic graph problems, the theoretical algorithmic ideas have received very little attention from the practical perspective. %

In this work, we tackle the maximum (weight) independent set problem in a fully dynamic setting. For a given graph $G=(V,E)$ an \textit{independent set} is defined as a subset $\I\subseteq V$ of all vertices such that each pair of vertices in $\I$ are non adjacent. A \textit{maximum independent set} (MIS) describes an independent set with the highest possible cardinality. By transforming the graph $G$ into the complement graph $\overline{G}$ the maximum independent set problem results in the \textit{maximum clique} problem. However, for sparse graphs $G$, using a maximum clique solver is impractical as the complement $\overline{G}$ is very dense and therefore unlikely to fit in memory for all but the smallest instances. Another related problem is the \textit{minimum vertex cover} problem. Note that for a maximum independent set $\I$ of $G$,  $V\setminus \I$ is a minimum vertex cover. For a weighted graph $G=(V,E,\omega)$ with non-negative vertex weights given by a function $\omega:V \rightarrow \mathbb{R}_{\geq 0}$, the \textit{maximum weight independent set} (MWIS) problem is to find an independent set $\I$ with maximum weight $\omega(\I) = \sum_{v\in \I} \omega(v)$. The applications of the maximum weight independent set problem, as well as the related problems addressed above, can be used for solving different application problems such as long-haul vehicle routing~\cite{DBLP:journals/corr/abs-2203-15805}, the winner determination problem~\cite{wu2015solving} or prediction of structural and functional sites in proteins~\cite{mascia2010predicting}. 

As a detailed example, consider an application of MWIS for (dynamic) map labeling, where displaying non-overlapping labels throughout dynamic map operations such as zooming and rotating~\cite{gemsa2016evaluation} or while tracking a physical movement of a user or set of moving entities~\cite{barth2016temporal} is of high interest in many applications. In the underlying map labeling problem, the labels are represented by vertices in a graph, weighted by importance. Each pair of vertices is connected by an edge if the two corresponding labels would overlap. 
In this graph, a MWIS describes a high quality set of labels, with regard to their importance level, that can be visualized without any overlap. 

Since these problems are NP-hard~\cite{DBLP:books/fm/GareyJ79}, heuristic
algorithms are used in practice to efficiently compute solutions of high quality on \textit{large} graphs~\cite{andrade-2012,grosso2008simple,DBLP:conf/gecco/GrossmannL0S23}. 
Depending on the definition of the neighborhood, local search algorithms are able to explore local solution spaces very effectively. 
However, local search algorithms are also prone to get stuck in local optima. 
As with many other heuristics, results can be improved if several repeated runs are made with some measures taken to diversify
the search. 
Still, even a large number of repeated executions can only scratch the surface of the huge space of possible independent sets for large-scale~data~sets. 

Traditional branch-and-bound methods~\cite{segundo-recoloring,segundo-bitboard-2011,tomita-recoloring,DBLP:conf/alenex/GellnerLSSZ21,DBLP:conf/alenex/Lamm0SWZ19,DBLP:conf/siamcsc/HespeL0S20} may often solve
small graphs with hundreds to thousands of vertices in practice, and medium-sized instances can be solved exactly in practice using reduction rules to reduce the graph. In particular, it has been observed that if data reductions work very well, then the instance is likely to be solved. If data reductions do not work very well, \ie the size of the reduced graph is large, then the instance can often not be solved.
Even though new algorithms such as the struction algorithm~\cite{DBLP:conf/alenex/GellnerLSSZ21} already manage to solve a lot of large instances, some remain unsolved.
In order to explore the global solution space extensively, more sophisticated metaheuristics, such as GRASP~\cite{DBLP:journals/corr/abs-2203-15805}, iterated local search~\cite{andrade-2012,hybrid-ils-2018} or memetic algorithms~\cite{DBLP:conf/gecco/GrossmannL0S23} have been used.

While quite a large amount of engineering work has been devoted to the computation of independent sets/vertex covers in static graphs, the amount of engineering work for the dynamic independent set problem is very limited with only three results by Zheng \etal~\cite{DBLP:conf/icde/ZhengWYC018,DBLP:conf/icde/ZhengPCY19} as well as Gao~\etal~\cite{gao2022dynamic} and one result by Bhore~\etal~\cite{DBLP:conf/esa/Bhore0N20} that is specialized \hbox{to~independent~rectangles}.
In this work, we extend the set of solvers for the maximum (weight) independent set problem in the dynamic model by introducing a technique that we call \textit{optimal neighborhood exploration}.

\paragraph*{Our Results.}
The \emph{basic idea} of our algorithms is as follows. Assume we are given a subset $H$ of the vertices $V$ in our graph $G$, along with an initial independent set $\mathcal{I}$. If we extend $H$ by including adjacent nodes that are also in the initial independent set, i.e., $H' \leftarrow H \cup (N(H) \cap \mathcal{I})$, then the intersection $(N(H') \setminus H') \cap \mathcal{I}$ will be empty. This ensures that we can replace the independent set nodes in $H'$ with any independent set from the subgraph induced by $H'$, denoted as $G[H']$. Consequently, to improve our solution, we can solve an independent set problem within the node-induced subgraph $G[H']$.
Based on this observation, we devise a local-search technique, called \term{\optNeighExplo} that builds independent induced subgraphs by exploring the neighborhood of a vertex locally up to a certain depth $d$. To solve the subproblems, we use recent state-of-the-art branch-and-reduce algorithms which are provided in the KaMIS library\footnote{\url{https://github.com/KarlsruheMIS}}. 
This yields a dynamic algorithm that can handle insertions and deletions.
To make the method feasible in practice, we propose various optimizations such as limiting the size of the subgraphs/subproblems, removing high degree vertices from the subproblems or performing expensive updates rarely.  
Note that in contrast to all theory of dynamic algorithms, our update operation has exponential worst-case time if we are interested in optimal solutions for the subgraphs. Still, our experiments show that the algorithms perform very well in practice. This opens a much wider discussion for dynamic algorithms with non-polynomial update time.
Lastly, we provide a simple greedy, fully-dynamic algorithm that provides good solutions \hbox{fast in practice.}

\section{Preliminaries} 

A graph $G=(V,E)$ is an undirected graph with 
${n=|V|}$ and ${m = |E|}$, where ${V =\{0,...,n-1\}}$. 
The neighborhood $N(v)$ of a vertex $v \in V$ is defined as ${N(v) = \{u \in V : (u,v) \in E\}}$. 
Additionally, $N[v]=N(v) \cup \{v\}$. 
The same sets are defined for the neighborhood $N(U)$ of a set of vertices ${U \subset V}$, \ie ${N(U) = \cup_{v \in U} N(v)\setminus U}$ and $N[U] = N(U) \cup U$. 
The degree of a vertex $\mathrm{deg}(v)$ is defined as the number of its neighbors $\mathrm{deg}(v)=|N(v)|$. 
The complement graph is defined as ${\overline{G}=(V,\overline{E})}$, where ${\overline{E}=\{(u,v): (u,v)\notin E\}}$ is the set of edges not present in $G$. 
A graph-sequence $\mathcal{G} = (G_0, \ldots, G_t)$ for $t\in\natnull$ is an \emph{edit-sequence of graphs} if 
for all $0 < i \leq t$ there exists exactly one edge $e\in E(G_{i})$ such that it is either inserted, \ie $G_{i} = G_{i-1} + e$, or deleted, \ie $G_{i} = G_{i-1} - e$, in update~$i$.%
A set $\I\subseteq V$ is called \textit{independent} if for all vertices $v,u \in \I$ there is no edge $(v,u)\in E$. 
An independent set is called \textit{maximal} if the set can not be extended by a vertex such that it is still independent. 
The \textit{maximum independent set problem} (MIS) is that of finding an independent set with maximum cardinality. 
The \textit{maximum weight independent set problem} (MWIS) is that of finding an independent set with maximum weight. 
The weight of an independent set $\I$ is defined as $\omega(\I) = \sum_{v \in \I}\omega(v)$ and $\alpha_w(G)$ describes the weight of a MWIS of the corresponding graph. 
Let $u\in V\setminus \I$ and $A\subset \I$.
We say $u$ is \term{tight} to $A$ if $A = N(u)\cap \I$.
The definition is motivated by the fact that $u$ can always join $\I$ if we remove $A$ from it.

The complement of an independent set is a vertex cover, \ie a subset ${C \subseteq V}$ such that every edge $e \in E$ is covered by at least one vertex $v \in C$. 
An edge is \textit{covered} if it is incident to one vertex in the set~$C$. 
The minimum vertex cover problem, defined as looking for a vertex cover with minimum cardinality, is thereby complementary to the maximum independent set problem. 
Another closely related concept are cliques. 
A \textit{clique} is a set $Q \subseteq V$ such that all vertices are pairwise adjacent. A clique in the complement graph $\overline{G}$ corresponds \hbox{to an independent set in the original graph~$G$.}

\section{Related Work}\label{sec:related_work}
\pagestyle{plain}
We give a short overview of existing work on both exact and heuristic procedures. For more details, we refer the reader to the recent survey on data reduction techniques~\cite{Abu-Khzam2022}.%

\subsection{Exact Methods.}

Exact algorithms usually compute optimal solutions by systematically exploring the solution space.
A frequently used paradigm in exact algorithms for combinatorial optimization problems is called \textit{branch-and-bound}~\cite{ostergaard2002fast,warren2006combinatorial}.
In case of the MWIS problem, these types of algorithms compute optimal solutions by case distinctions in which vertices are either included into the current solution or excluded from it, branching into two or more subproblems and resulting in a search tree.
Over the years, branch-and-bound methods have been improved by new branching schemes or better pruning methods using upper and lower bounds to exclude specific subtrees~\cite{balas1986finding, babel1994fast,li2017minimization}.
In particular, Warren and Hicks~\cite{warren2006combinatorial} proposed three branch-and-bound algorithms that combine the use of weighted clique covers and a branching scheme first introduced by Balas and Yu~\cite{balas1986finding}.
Their first approach extends the algorithm by Babel~\cite{babel1994fast} by using a more intricate data structure to improve its performance.
The second one is an adaptation of the algorithm of Balas and Yu, which uses a weighted clique heuristic that yields structurally similar results to the heuristic of Balas and Yu.
The last algorithm is a hybrid version that combines both algorithms and is able to compute optimal solutions on graphs \hbox{with hundreds of vertices.}

In recent years, reduction rules have frequently been added to branch-and-bound methods yielding so-called \textit{branch-and-reduce} algorithms~\cite{akiba-tcs-2016}. 
These algorithms are able to improve the worst-case runtime of branch-and-bound algorithms by applications of reduction rules to the current graph before each branching step.
For the unweighted case, a large number of branch-and-reduce
algorithms have been developed in the past. The currently best exact
solver~\cite{hespe2020wegotyoucovered}, which won the PACE challenge
2019~\cite{hespe2020wegotyoucovered, bogdan-pace, peaty-pace}, uses a portfolio of branch-and-reduce/bound solvers for the complementary  problems. Recently, novel branching strategies have been presented in~\cite{DBLP:conf/wea/HespeLS21} to further improve both branch-and-bound as well as branch-and-reduce approaches.

However, for a long time, virtually no weighted reduction rules were known, which is why hardly any branch-and-reduce algorithms exist for the MWIS problem. The first branch-and-reduce algorithm for the weighted case was presented by Lamm~\etal~\cite{lamm-2019}.
The authors first introduce two meta-reductions called neighborhood removal and neighborhood folding, from which they derive a new set of weighted reduction rules.
On this foundation a branch-and-reduce algorithm is developed using pruning with weighted clique covers similar to the approach by Warren and Hicks~\cite{warren2006combinatorial} for upper bounds and an adapted version of the ARW local search~\cite{andrade-2012} for lower bounds.

This algorithm was then extended by Gellner~\etal~\cite{DBLP:conf/alenex/GellnerLSSZ21} to utilize different variants of the struction, originally introduced by Ebenegger~\etal~\cite{ebenegger1984pseudo} and later improved by Alexe~\etal~\cite{alexe2003struction}. In contrast to previous reduction rules, these were not necessarily decreasing the graph size, but rather transforming the graph which later can lead to even further reduction possibilities. Those variants were integrated into the framework of Lamm~\etal~\cite{lamm-2019} in the preprocessing as well as in the reduce step.
The experimental evaluation shows that this algorithm can solve a large set of real-world instances and outperforms the branch-and-reduce algorithm by Lamm~\etal~\cite{lamm-2019}, as well as different state-of-the-art heuristic approaches such as the algorithm {\hils} presented by Nogueira~\cite{hybrid-ils-2018} as well as two other local search algorithms \textsf{DynWVC1} and \textsf{DynWVC2} by Cai~\etal~\cite{cai-dynwvc}. Recently, Xiao~\etal~\cite{xiao2021efficient} present further data reductions for the weighted case as well as a simple exact algorithm based on these data reduction rules. Furthermore, in~\cite{DBLP:conf/icde/ZhengGPY20} a new reduction-and-branching algorithm was introduced using two new reduction rules.
Not long ago Huang~\etal~\cite{DBLP:conf/cocoon/HuangXC21} also presented a branch-and-bound algorithm using reduction rules working especially well on sparse graphs. In their work they additionally undertake a detailed analysis for the running time bound on special graphs. With the measure-and-conquer technique they can show that for cubic graphs the running time of their algorithm is~$\mathcal{O}^*(\numprint{1.1443}^n)$ which is improving previous time bounds for this problem using polynomial space complexity and for graphs of average degree three.
Figiel~\etal~\cite{DBLP:conf/esa/FigielFNN22}~introduced a new idea added to the state-of-the-art way of applying reductions. They propose to not only performing reductions, but also the possibility of undoing them during the reduction process. As they showed in their paper for the unweighted independent set problem, this can lead to new possibilities to apply further reductions and finally to smaller reduced graphs.

Finally, there are exact procedures which are either based on other extension of the branch-and-bound paradigm, e.g.~\cite{rebennack2011branch,warrier2005branch,warrier2007branch}, or on the reformulation into other $\mathcal{NP}$-complete problems, for which a variety of solvers already exist.
For instance, Xu~\etal~\cite{xu2016new} developed an algorithm called \textsf{SBMS}, which calculates an optimal solution for a given MWVC instance by solving a series of SAT instances. Also for the MWVC problem a new exact algorithm using the branch-and-bound idea combined with data reduction rules were recently presented \cite{DBLP:journals/corr/abs-1903-05948}.
We additionally note that there are several recent works on the complementary maximum weighted clique problem that are able to handle large real-world networks~\cite{fang2016exact,jiang2017exact,held2012maximum}.
However, using these solvers for the maximum weight independent set problem requires computing complement graphs.
Since large real-world networks are often very sparse, processing their complements quickly becomes infeasible due to their memory~requirement.

\subsection{Heuristic Methods.}

A widely used heuristic approach is local search, which usually computes an initial solution and then tries to improve it by simple insertion, removal or swap operations. 
Although in theory local search generally offers no guarantees for the solution's quality, in practice they find high quality solutions significantly faster than exact procedures.

For unweighted graphs, the iterated local search (ARW) by Andrade~\etal~\cite{andrade-2012}, is a very successful heuristic.
It is based on so-called $(1,2)$-swaps which remove one vertex from the solution and add two new vertices to it, thus improving the current solution by one.
Their algorithm uses special data structures which find such a $(1,2)$-swap in linear time in the number of edges or prove that none exists.
Their algorithm is able to find (near-)optimal solutions for small to medium-size instances in milliseconds, but struggles on massive instances with millions of vertices and edges.

The hybrid iterated local search (\hils) by Nogueira~\etal~\cite{hybrid-ils-2018} adapts the ARW algorithm for weighted graphs.
In addition to weighted $(1,2)$-swaps, it also uses $(\omega,1)$-swaps that add one vertex $v$ into the current solution and exclude its $\omega$ neighbors. 
These two types of neighborhoods are explored separately using variable neighborhood descent (VND).

Two other local searches, DynWVC1 and DynWVC2, for the equivalent minimum weight vertex cover problem are presented by Cai~\etal~\cite{cai-dynwvc}.
Their algorithms extend the existing FastWVC heuristic~\cite{li2017efficient} by dynamic selection strategies for vertices to be removed from the current solution.
In practice, DynWVC1 outperforms previous MWVC heuristics on map labeling instances and large scale networks, and DynWVC2 provides further improvements on large scale networks but performs worse on map labeling instances.

Li~\etal~\cite{li2019numwvc} presented a local search algorithm for the minimum weight vertex cover (MWVC) problem, which is complementary to the MWIS problem. Their algorithm applies reduction rules during the construction phase of the initial solution.
Furthermore, they adapt the configuration checking approach~\cite{cai2011local} to the MWVC problem which is used to reduce cycling, \ie returning to a solution that has been visited recently.
Finally, they develop a technique called self-adaptive-vertex-removing, which dynamically adjusts the number of removed vertices per iteration.
Experiments show that their algorithm outperforms state-of-the-art approaches  graphs of up to millions of vertices.

Recently, a new hybrid method was introduced by Langedal~\etal~\cite{langedal2022efficient} to also solve the MWVC problem. For this approach they combined elements from exact methods with local search, data reductions and graph neural networks. In their experiments they achieve definite improvements compared to \textsf{DynWVC2} and the {\hils} algorithm in both solution quality and running~time.

With \textsf{EvoMIS}, Lamm~\etal~\cite{lamm2015graph} presented an evolutionary approach to tackle the maximum independent set problem. The key feature of their algorithm is to use graph partitioning to come up with natural combine operations, where whole blocks of solutions to the MIS problem can be exchanged easily. To these combine operations also local search algorithms were added to improve the solutions further.
Combining the branch-and-reduce approach with the evolutionary algorithm \textsf{EvoMIS}, a reduction evolution algorithm \textsf{ReduMIS} was presented by Lamm~\etal~\cite{redumis-2017}. In their experiments, \textsf{ReduMIS} outperformed the local search \textsf{ARW} as well as the pure evolutionary approach \textsf{EvoMIS}. Later, a memetic approach for maximum weight independent set problem has been proposed by Großmann \etal~\cite{DBLP:conf/gecco/GrossmannL0S23}.%

Another reduction based heuristic called {\htwis} was presented recently by Gu~\etal~\cite{gu2021towards}. 
The repeatedly apply their reductions exhaustively and then choose one vertex by a tie-breaking policy to add to the solution. Now this vertex as well as its neighbors can be removed from the graph and the reductions can be applied again.
Their experiments prove a significant improvement in running time as well as computed weights of the MWIS compared to the state-of-the-art~solvers. 

Recently, a new metaheuristic for the greedy randomized adaptive search procedure (GRASP) framework by Resende et al.~\cite{resende2016optimization} was introduced by Goldberg~\etal~\cite{DBLP:journals/corr/abs-2203-15805} in particular for vehicle routing instances~\cite{dong2021new}. With their algorithm \textsc{METAMIS} they developed a new local search algorithm combining a wide range of simple local search operations with a new variant of path-relinking to escape local optima.
Besides known swap operations as $(\omega,1)$-, $(1,\omega)$- and $(2,\omega)$-swaps, they present a new local-search procedure called alternating augmenting path (AAP) moves.
An AAP is defined as path such that solution vertices and non-solution vertices along the path alternate in the order they appear and flipping the vertices,~\ie exchanging the vertices between the independent set and the complement, must maintain an independent set.
The move is accepted if the solution weight is improved by flipping the path.
In their experiments they outperform {\hils} algorithm on a wide range of instances both in time and solution~quality. 

\subsection{Dynamic Independent Sets.} Recently, there has been a survey about fully dynamic graph algorithms by Hanauer~\etal~\cite{DBLP:journals/corr/abs-2102-11169} which also covers the independent set problem. We follow their description closely: As computing the size of an MIS is NP-hard, all dynamic algorithms of independent set study the maximal independent set problem. 
In a sequence of papers~\cite{DBLP:conf/stoc/AssadiOSS18,DBLP:journals/corr/abs-1804-01823,DBLP:conf/soda/AssadiOSS19,DBLP:conf/focs/ChechikZ19,DBLP:conf/focs/BehnezhadDHSS19} the running time for 
the maximal independent set problem was reduced to $O(\log^4 n)$ expected worst-case update time. All these algorithms actually maintain a maximal independent set. A query can either return the size of that set in constant time or output the whole set in time linear in its size.

While quite a large amount of engineering work has been devoted to the computation of independent sets/vertex covers in static graphs (see above), the amount of engineering work for the dynamic independent set problem is very limited.
Zheng \etal~\cite{DBLP:conf/icde/ZhengWYC018} presented a heuristic fully dynamic algorithm and proposed a lazy search algorithm to improve the size of the maintained independent set.
A year later, Zheng \etal~\cite{DBLP:conf/icde/ZhengPCY19} improved the result such that the algorithm is less sensitive to the quality of the initial solution used for the evolving MIS. In their algorithm, the authors used two well known data reduction rules, degree one and degree two vertex reduction,  that are frequently used in the static case.   Moreover, the authors can handle batch updates. Bhore \etal~\cite{DBLP:conf/esa/Bhore0N20} focused on the special case of MIS for independent rectangles which is frequently used in map labeling applications. The authors presented a deterministic algorithm for maintaining a MIS of a dynamic set of uniform rectangles with amortized sub-logarithmic update time. Moreover, the authors evaluated their approach using extensive experiments.

Recently, Gao~\etal~\cite{gao2022dynamic} published a dynamic approximation algorithm for the maximum independent set problem, which relies on swapping solution and non-solution nodes. They show that their algorithm maintains an $(\frac{\Delta}{2} +1)$-approximate solution over dynamic graphs where $\Delta$ is the maximum degree of the graph.

\begin{figure}
\centering
\includegraphics[width=4cm]{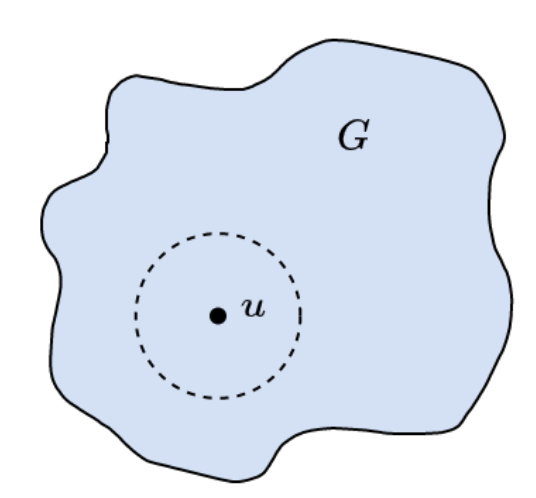}
\caption{Starting from a node $u$, the algorithms find a locally independent subgraph $G[H]$ by exploring the neighborhood up to a certain depth and afterwards adding adjacent nodes from the current solution to the subgraph. The subgraph $G[H]$ is then solved by a solver of our choice.}
\vspace*{-.5cm}
\label{fig:induced_subgraph}
\end{figure}

\section{Optimum Neighborhood Exploration}\label{subsec: one}
Before we explain our dynamic algorithm, we explain the core concept of optimum neighborhood exploration.
The core idea that we use to improve a given weighted independent set (or handle updates in a dynamic graph) is as follows.
We first build an induced subgraph~$G[H]$  such that it has the property that its nodes $H$ in the original graph are not adjacent to independent set nodes outside of $H$, \ie $(N_G(H)\backslash H) \cap \mathcal{I} = \emptyset$. This implies that we can exchange the independent set nodes of $\mathcal{I}\cap H$ in~$G$ by any independent set of the node induced subgraph. Hence, in order to improve a solution, we solve the maximum weight/cardinality independent set problem in the node induced subgraph~$G[H]$.  The running time of the algorithm depends crucially on the size of the induced subgraph.  Hence, for our dynamic approach we limit the size of the subgraph.

\subsection{Building the Induced Subgraph.}
We now explain how we find the induced subgraphs for our improvement algorithm. We start with the basic idea. 
Assume that a maximal (weight) independent set $\IS$ of~$G$ is given. We build the subgraph $G[H]$ starting around a seed vertex~$u\in V$.
To construct the subgraph, we perform a breadth first search (BFS) of depth $d$ for a given parameter $d > 0$.
All nodes reached by the BFS are added to a set~$H$. Afterwards, we add all nodes that are adjacent to nodes in $H$ and in the given independent set $\mathcal{I}$. More precisely, we set $H \leftarrow H \cup (N_G(H) \cap \mathcal{I})$. This ensures that $(N_G(H) \backslash H)\ \cap\  \mathcal{I} = \emptyset$ and therefore we can exchange the independent set nodes of $\mathcal{I}$ in $H$ by any independent set of the node induced subgraph~$G[H]$. 

\begin{figure}
	\centering
	\input{figures/non_maximal.tex}
	\caption{Induced subgraph with BFS of depth 1 starting at $u$. Current solution vertices  are orange. The set $H = N[u]\cup \{v\}$. The new optimal solution on the subgraph is green. When changing to this solution, the independent set is not maximal.}\label{fig:non_maximal}
        \vspace*{-.25cm}
\end{figure}
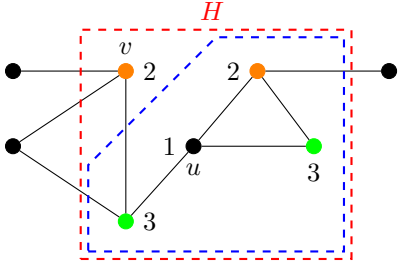

Note that if we only consider one single local problem in an induced subgraph around $u$, it is possible that the improved independent set in $G$ is not maximal even if the solution in $G[H]$ has been optimal, see Figure~\ref{fig:non_maximal} for an example. We fix this by adding nodes in $N_G(H) \backslash H$ for which all adjacent independent set nodes are in $H$, \ie we add the $H$ tight nodes from $N(H) \backslash H$ \hbox{to our subproblem.}

\subsubsection{Limiting Subproblem Size.} 
The basic idea can yield large subgraphs $G[H]$ if the expansion of the graph is high. 
Consequently, we not only limit the depth of the BFS but also introduce a second parameter, $\nu_{\text{max}} > 0$, to cap the total number of vertices in $H$. 
In order to limit the size of the induced subgraph we perform a modified breadth first search. We start the BFS at $u$ and put it into the BFS queue~$Q$. Whenever a node $v$ is reached by the BFS we know the size of the current set $H$ and its extended set via bookkeeping. If adding~$v$ to $H$ would imply that the final problem size is larger than $\nu_\text{max}$ (when adding respective independent set nodes and $H$ tight nodes of $N(H) \backslash H$), we do not add this vertex to $H$ and also do not continue the BFS at that node, \ie $v$ is not added to the BFS queue $Q$. The BFS continues until all nodes have been processed until at most depth~$d$ or the queue $Q$ is empty.

\begin{algorithm}[t!]
    \caption{\textsc{DegGreedy} Algorithm}\label{algo:deggreedy}
    \begin{algorithmic}
        \Procedure{insertion}{$u$,$v$}
            \State \textsc{Adj}[$u$] $:=$ \textsc{Adj}[$u$] $\cup \, \{v\}$
            \State \textsc{Adj}[$v$] $:=$ \textsc{Adj}[$v$] $\cup\, \{u\}$

            \If{$u \in \textsc{IS}$ \textbf{and} $v \in \textsc{IS}$}
                \State $\phi_v \gets w(v)/w(N(v))$
                \State $\phi_u \gets w(u)/w(N(u))$
                \State $\mathcal{R} \gets u$\Comment{$\mathcal{R}$ stores vertex to be removed}

                \If{$\phi_v < \phi_u$}
                    \State  $\mathcal{R}\gets v$                 
                \ElsIf{$\phi_u = \phi_v$}
                    \State $\mathcal{R} \gets$ \textsc{Random}($u$ or $v$)
                \EndIf
                
                \State $\mathcal{I}$ $:=$ $\mathcal{I}$ $\setminus \{\mathcal{R}\}$

                \For{$v \in N(\mathcal{R})$}
                    \If{\textsc{CanBeIndependent}($v$)}
                        \State $\mathcal{I}$ $:=$ $\mathcal{I}$ $\cup \, \{v\}$
                    \EndIf
                \EndFor
            \EndIf
        \EndProcedure

        \Procedure{deletion}{$u$,$v$}
            \State \textsc{Adj}[$u$] $:=$ \textsc{Adj}[$u$] $\setminus \{v\}$
            \State \textsc{Adj}[$v$] $:=$ \textsc{Adj}[$v$] $\setminus \{u\}$

            \If{$u \notin \mathcal{I}$}
                \If{\textsc{CanBeIndependent}($u$)}
                    \State $\mathcal{I}$ $:=$ $\mathcal{I}$ $\cup \, \{u\}$
                \EndIf
            \EndIf

            \If{$v \notin \mathcal{I}$}
                \If{\textsc{CanBeIndependent}($v$)}
                    \State $\mathcal{I}$ $:=$ $\mathcal{I}$ $\cup \, \{v\}$
                \EndIf
            \EndIf
        \EndProcedure

        \Procedure{CanBeIndependent}{$u$}
            \State \Return $N(u) \cap \mathcal{I} = \emptyset$
        \EndProcedure

    \end{algorithmic}
\end{algorithm}
\subsubsection{Pruning Large Degree Vertices (Pinching).}
\label{subsec:pruning}
Large degree vertices in sparse graphs are often not part of optimum solutions (or vertices where $\omega(N(v))/\omega(v)$ is high for the weighted case).
To address this, we implement a pruning strategy that selectively removes high degree vertices from the subproblem~$G[H]$ before solving it. By excluding these vertices, we reduce the complexity of the subproblem, allowing the solver to concentrate on more promising candidate vertices with lower degrees. This approach not only speeds up the solution process but also helps maintain a more manageable subproblem size, ensuring the solver can operate within practical time limits. This technique is frequently employed in static (heuristic) algorithms that use data reductions, effectively breaking up the reduction space after exhaustive application of data reductions, e.g.~\cite{DBLP:conf/gecco/GrossmannL0S23}. Here, we look at the current subgraph~$G[H]$ and first find the independent set vertex in~$G[H]$ with the largest degree. Let the degree of that vertex be denoted by $\Delta_{\mathcal{I},H}$. Then, before solving $G[H]$, we remove vertices from $H \setminus \mathcal{I}$, and thus from the subproblem we want to solve, having a degree larger than  $\delta \cdot \Delta_{\mathcal{I},H}$ where $\delta > 1$ is a tuning parameter. We use $\delta = 1.25$, however, the algorithm is not too sensitive about the precise \hbox{choice of the parameter.}

\subsection{Solving Local Induced Subgraphs.}
In principle, it is possible to use any (exact) solver which aims to solve the local MWIS problem optimally or heuristically.
We choose the \baR solver from KaMIS (\algoname{KaMIS~BnR}) by Lamm~\etal~\cite{DBLP:conf/alenex/Lamm0SWZ19} since it is state-of-the-art among the exact algorithms and it can solve weighted as well as cardinality independent set problems.
In general, the algorithm applies a wide range of data reductions to the input instance exhaustively and afterwards runs a sophisticated branch-and-reduce algorithm.
It is capable to solve many (large) static instances optimally while outperforming other approaches.
Even compared to \ils approaches like HILS it can often compete in run-time, \eg for OSM instances (street networks), as the experiments by Lamm \etal~\cite{DBLP:conf/alenex/Lamm0SWZ19} show.
We run  \algoname{KaMIS~BnR} with a time-limit $t_{\text{BnR}}$. If the algorithm can not find an optimal solution within the time limit, it returns the best weighted independent set~$\IS_H$ for~$G[H]$ that it has found.
Since every independent set for~$G[H]$ yields a modified and feasible independent set in~$G$, we update the old solution~$\IS$ in~$G$ if~$\IS_H$ improves solution quality, \ie if $\wMap(\IS_H)>\wMap(\IS\cap H)$. %

\begin{algorithm}[t!]
    \caption{DynamicONE Algorithm}\label{algo:dynamicone}
    \begin{algorithmic}
        \Procedure{insertion}{$u$,$v$}
            \State \textsc{Adj}[$u$] $:=$ \textsc{Adj}[$u$] $\cup \{v\}$
            \State \textsc{Adj}[$v$] $:=$ \textsc{Adj}[$v$] $\cup \{u\}$

            \State \textsc{DegGreedy::insertion}($u$, $v$)

            \If{\textsc{DegGreedy} cannot add a vertex to $\mathcal{I}$}
                \State $H \gets$ \textsc{BFS-SEARCH}($u$, $v$, $d$, $\nu_{\text{max}}$)
                \State \textsc{PruneLargeDegreeVertices}($H$)
                \State $H \gets H \cup (N_G(H) \cap \mathcal{I})$
                \If{\textsc{SolveIS}($G[H]$, $\mathcal{I} \cap H$)}
                    \State \textsc{UpdateSolution}()
                \EndIf
            \EndIf
        \EndProcedure

        \Procedure{deletion}{$u$,$v$}
            \State \textsc{Adj}[$u$] $:=$ \textsc{Adj}[$u$] $\setminus \{v\}$
            \State \textsc{Adj}[$v$] $:=$ \textsc{Adj}[$v$] $\setminus \{u\}$

            \State \textsc{DegGreedy::deletion}($u$, $v$)

            \If{\textsc{DegGreedy} cannot add a vertex to $\mathcal{I}$}
                \State $H \gets$ \textsc{BFS-SEARCH}($u$, $v$, $d$, $\nu_{\text{max}}$)
                \State \textsc{PruneLargeDegreeVertices}($H$)
                \State $H \gets H \cup (N_G(H) \cap \mathcal{I})$
                \If{\textsc{SolveIS}($G[H]$)}
                    \State \textsc{UpdateSolution}()
                \EndIf            \EndIf
        \EndProcedure

        \Procedure{PruneLargeDegreeVertices}{$H$}
            \For{each $v \in H$}
                \If{$d(v) > \delta\cdot\Delta_{\mathcal{I},H}$ and $v \not \in \mathcal{I}$}
                    \State Remove $v$ from $H$
                \EndIf
            \EndFor
        \EndProcedure

        \Procedure{SolveIS}{$G_H$, $\mathcal{I}_\mathcal{H}$}
            \State Solve (weighted) independent set problem in $G_H$
            \State \Return true if solution improved, false otherwise
        \EndProcedure

        \Procedure{UpdateSolution}{}
            \State Update $\mathcal{I}$ in $G$ with solution of $G[H]$
        \EndProcedure
    \end{algorithmic}
\end{algorithm}

\section{Fully Dynamic Maximum Weight and Cardinality Independent Sets}\label{subsec:fullydynamic}
We will now leverage the core concept of optimum neighborhood exploration described above to define a fully dynamic algorithm for the maximum weight and the maximum cardinality independent set problem. Next, we will explain our approach to handling various update operations. Before we explain dynamic optimum neighborhood exploration, we give simple dynamic and fast algorithms. Later, we combine these fast algorithms with our (more expensive) dynamic optimum neighborhood exploration to obtain a fast and high-quality fully dynamic algorithm.
\subsection{Fast and Simple Greedy Updates.} 
When a new edge $\{u,v\}$ is \emph{inserted}, we first check if the newly created edge induces a conflict. Specifically, we verify whether both \(u\) and \(v\) are currently part of the independent set. If they are, the current solution is no longer independent. To resolve this, we remove the node which minimizes $w(v)/w(N(v))$. In case of a tie, we remove a random vertex from the solution. Note that in the cardinality case, this reduces to removing the vertex with the larger degree from the current solution. The intuition here is that we want to remove the node that blocks the most other vertices/weight in the graph. Let the vertex that we removed from the solution be $u$. Since $u$ is no longer an independent set node, we check if its neighbors can be added to the solution. Checking each neighbor takes $O(\Delta)$ time, thus the overall update takes time $O(\Delta^2)$. Note that by using proper data structures, \ie storing and updating for each vertex the number of adjacent independent set vertices, one can reduce the update time to $O(\Delta)$. If an edge $\{u,v\}$ is \emph{deleted} from the graph, we check in $O(\Delta)$ time if $u$ and $v$ can be added to the solution (if $u$ or $v$ are not already part of the solution). We call this fast and simple algorithm \textsc{DegGreedy}. Pseudocode for the algorithm \hbox{is given in Algorithm~\ref{algo:deggreedy}.}
\subsection{Dynamic Optimum Neighborhood Exploration.} We now explain our approach to performing dynamic optimum neighborhood exploration. In general, we integrate this with the fast and simple greedy updates to optimize running time. We call this technique \emph{pruning updates}. Initially, we execute the aforementioned fast and simple greedy algorithm. If this algorithm can add a vertex to the solution after updating the graph data structure, no further action is taken.  However, if it cannot, we then use the more resource-intensive dynamic optimum neighborhood exploration to attempt a local improvement of the solution. Pruning the expensive updates using the fast and simple update algorithm improves the running time of the overall algorithm significantly. We refer to this algorithm as \textsc{DynamicONE}. Pseudocode for the algorithm is given in Algorithm~\ref{algo:dynamicone}.

\paragraph{Edge Insertion.}
We construct the subgraph $H$ by performing a breadth-first search using the two parameters $d$ and $\nu_\text{max}$. We initialize the BFS with both nodes $u$ and $v$. We then prune the large degree vertices of the subgraph (see Section~\ref{subsec:pruning}). Afterwards, we solve the (weighted) independent set problem in the induced subgraph and update the solution accordingly if we found an improvement. 

\paragraph{Edge Deletion.}
Deleting an edge can not lead to any conflict between independent set vertices. Hence, the current independent set does not need any type of fixing.
Hence, after running the fast and greedy algorithm unsuccessfully, we construct the subgraph $H$ based on the two parameters $d$ and $\nu_{\text{max}}$ to obtain the subgraph $H$. As in the insertion case, we initialize the breadth-first search with both nodes $u$ and $v$. As before, we prune the large degree vertices of the subgraph. We then solve the (weighted) independent set problem in the induced subgraph and update the solution accordingly. 

\paragraph{Miscellaneous.}
To save running time, we limit the time of the local solver to 10s and define \emph{rare updates}. Rare updates only perform expensive updates (solving subgraphs) every $x=3$ updates. In the other two cases, only greedy updates are performed. Note that the parameter $x$ is a trade-off parameter between running time and solution quality. A smaller value of $x$ will lead to better solution quality but also to a higher running time. Due to space constraints, we do not present results of the experiments with the rare updates for different values of $x$.
Other update operations such as node insertion (or weight update) and deletion of a node $u$ can be done by mapping those operations to edge insertion and deletions. 
However, this can yield a large amount of subproblems that need to be solved, for example if the degree of the inserted vertex is large. One can improve node insertion (or weight update) by solving just one subgraph $H$ starting from the seed node $u$ (with its edges being already inserted in the graph). 
When deleting a node $u$, we can initialize the dynamic BFS with the former neighbors of $N(u)$ to obtain a subgraph and hence a subproblem.

\section{Experimental Evaluation}
\paragraph{Methodology.}
We implemented our algorithm using C++17. The code is compiled using g++ version 11.4.0 and full optimizations turned on (-O3).
In the unweighted dynamic case, we compare against 
the dynamic implementations of Zheng \etal~\cite{DBLP:conf/icde/ZhengPCY19} (\texttt{DgOracleTwo}) as well as Gao~\etal~\cite{gao2022dynamic} (\texttt{DyTwoSwap}).
These algorithms are also implemented in C++ and compiled with the same compiler and optimization settings as our algorithm.
In general, we compare the results of the algorithms after all \hbox{updates have been performed.}
We generally run each algorithm/configuration with ten different seeds. 
We have used a machine with an  AMD EPYC 9754 128-Core CPU running at 2.25GHz with 256MB L3 Cache and 768GB of main memory. 
It runs Ubuntu GNU/Linux 22.04 \hbox{and Linux kernel version 5.15.0-102}. 

\paragraph{Data Sets.}
We compare ourselves on a wide range of dynamics instances. These instances have been collected 
from various resources~\cite{benchmarksfornetworksanalysis,UFsparsematrixcollection,snap,DBLP:conf/www/Kunegis13,konect:unlink,DBLP:journals/corr/abs-2102-11169}.
Table~\ref{dyninstances} summarizes the main properties of the dynamic benchmark set.
Our benchmark set includes a number of graphs from numeric simulations, road networks as well as complex networks/social networks.
Our set includes static graphs as well as real dynamic graphs. In case of instances that have originally been static, we create dynamic instances by starting with an empty graph and insert all edges in order of their appearance of the static graph. 
As our algorithms only handles undirected simple graphs, we ignore possible edge directions in the input graphs, and we remove self-loops and parallel edges.
In general, for the cardinality case we treat \hbox{all graphs as unweighted graphs.}

\subsection{Fully Dynamic Maximum Cardinality Independent Set Algorithms.}
We now present an evaluation of our algorithms for the maximum cardinality independent set problem. Due to space constraints, our focus will be on assessing the impact of the BFS search depth and the effectiveness of various algorithmic components in accelerating performance. Finally, we will compare our algorithm \hbox{against state-of-the-art competing algorithms.}

\textbf{Depth $d$.} We start with the depth of the BFS search, which is denoted by $d$. Note that this parameter is such that increasing the parameter increases the size of the local problems and thus will (likely) increase the running time of the algorithm, but also improve the quality of the solutions. Hence, we look at the performance of the parameter on all instances. Note that the same parameter is used by the respective algorithm on all instances and that there is no instance specific tuning of the parameter.  %

\begin{figure}[t]
\centering
\vspace*{-1cm}
\hspace*{0.25cm}\includegraphics[width=0.5\textwidth]{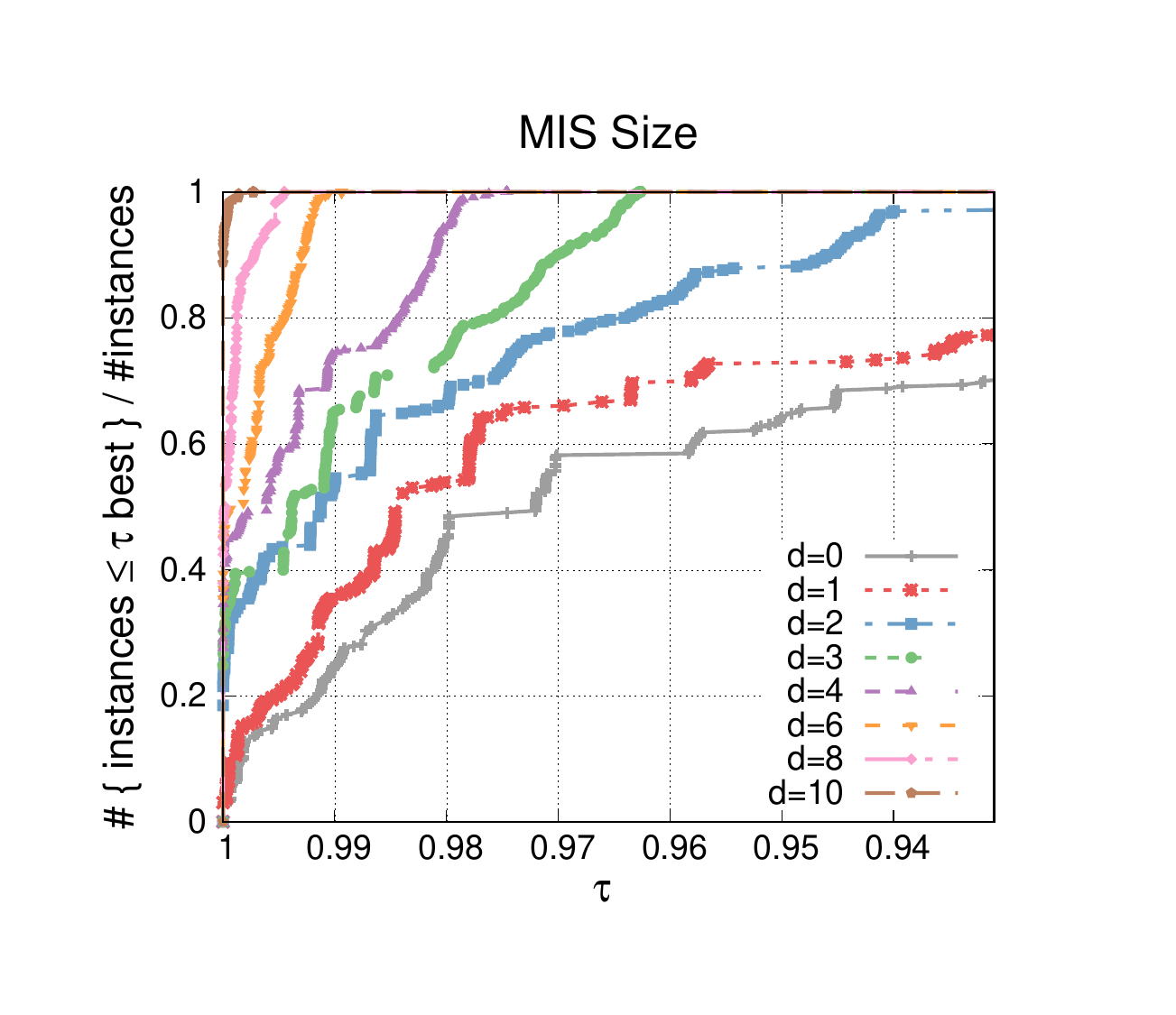} \\
                \vspace*{-.5cm}
\includegraphics[width=0.48\textwidth]{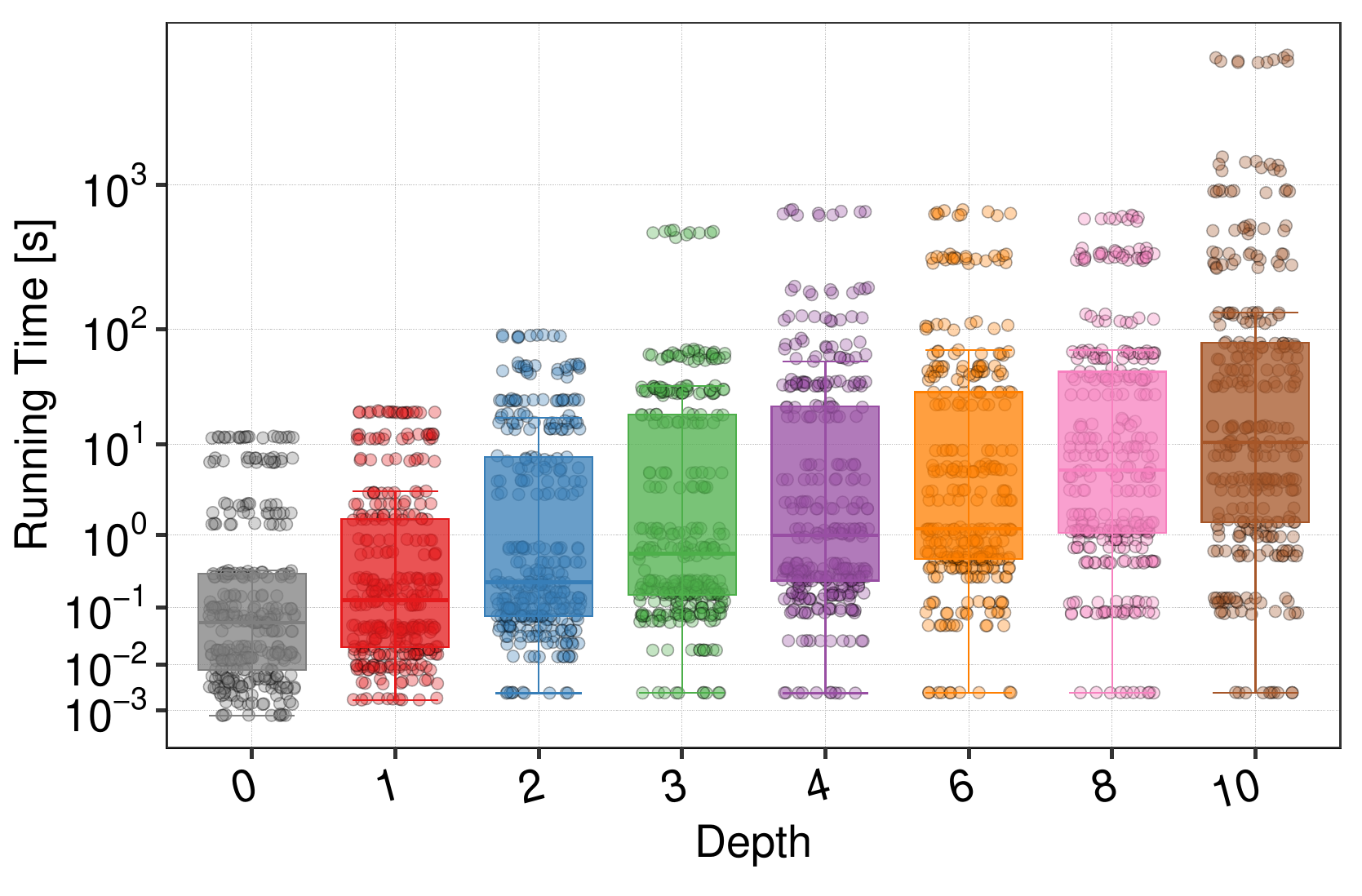}
\caption{Performance profiles (top) and running times in total for all updates (bottom) for various depths $d$ with $\nu_\text{max}=2500$.}
\label{fig:performancedepth}
\end{figure}

In this experiment, we set $\nu_\text{max} = 2500$ and use $d \in \{0, 1, 2, 3, 4, 6, 8, 10\}$. Figure~\ref{fig:performancedepth} provides a performance profile for solution quality and a running time box plot for the different algorithms as a function of depth $d$. The algorithms behave as expected, with the quality of the solution improving as the depth of the BFS search increases. However, this improvement comes with a significant increase in running time. For example, increasing the depth from $d=0$ to $10$ results in an average solution improvement of 6\%, with the largest observed improvement being 22\% on the \texttt{wing} instance. However, this depth also makes the algorithm 181 times slower. The plots further show that initial improvements in quality are more substantial, such as when increasing depth from $d=0$ to $4$. At $d=4$, our algorithm yields solutions that are 5\% better than the algorithm using $d=0$ on average, and it is 25 times slower. Thus in applications this parameter can be effectively used as a parameter to control the quality/size of solutions.  However, further techniques (as will follow) are required to speed up the algorithm. 

\textbf{Pruning Updates.} Pruned updates help avoid the expensive solving steps of the induced subgraph by performing a simpler and faster local search augmentation step, if successful. To evaluate the impact of the pruning updates on the algorithm's performance, we tested all values for $d$ used previously with pruned updates enabled. Overall, we observe that pruning updates significantly improve the algorithm's running time. The speed-up does not heavily depend on $d$, as the pruning step filters out unnecessary updates across all algorithm configurations. On average, pruning reduces the algorithm's running time by 21\%. Additionally, pruning slightly improves solution quality (by a very small margin), which may be attributed to minor variations in the solutions found by the algorithm throughout the various update steps. Hence, we conclude that pruning updates are a crucial component of our algorithm, as they significantly reduce the running time \hbox{without compromising solution quality.}

\textbf{Rare Updates.} With rare updates we introduce another technique to reduce the number of (expensive) updates that need to be solved. We tested the impact of rare updates on the algorithm's performance by varying the parameter running the same configurations as above (without pruning) as well as the configurations from above with rare updates enabled. The results show that rare updates can significantly reduce the algorithm's running time. Here, the impact varies for different values of $d$.
For small values of $d$, the running time without pruning and with pruning are closer, but as $d$ increases, the benefit of using rare updates becomes more pronounced. For instance, when $d=1$, the running time improved by 35\% with rare updates (but size of solutions also decreased by 17\%). As $d$ increases to 10, the running time is improved by a factor 3.05 while solutions are worse by 0.6\% on average. 
The small decrease in solution quality is due to the fact that ignored updates likely take part in a later update as the local subgraphs can be quite large for large values of $d$.
In summary, rare updates prove to be a effective technique for reducing running time, especially for larger values of $d$, with small impact on the solution quality. Still, if solution quality is paramount, this option may not be useful. Note that in principle different values of $x$ could yield different trade-offs. However, due to space constraints we do not explore this here. %

\textbf{Pinching.} Pinching removes nodes from the local subgraph that are unlikely to be in an independent set and hence reduces the size of the local subgraph. 
In addition to reducing the size, removing these nodes also makes data reductions more effective. We tested the impact of pinching on the algorithm's performance by varying the parameter running the same configurations as above (without pruning and rare updates) as well as the configurations from above with pinching enabled. The results show that pinching can significantly reduce the  running time of the algorithm. As expected, for small values of $d$ pinching does not have a high impact, neither on solution quality nor on running time. However, for larger values of $d\geq 3$ solution quality is not impacted ($<0.001$\% on average) while running time is reduced by 12\%, 15\%\, 18\%, 22\% and 16\% for $d=3,4,6,8,10$ respectively. Hence, we conclude that pinching is a useful technique for reducing the running time of the algorithm without compromising solution quality for values of $d \geq 3$. %

\textbf{Summary.} Based on these observations, we define two version of our algorithm. \texttt{DynamicOneStrong} is the version of our algorithm that uses a depth of $d=10$ and having pinching and pruning enabled. \texttt{DynamicOneFast} uses $d=10$ with all of the components from above (pinching, pruning, rare updates) enabled, additionally we decrease the maximum allowed local subproblem size and use $\nu_\text{max}=200$ to further speed up our algorithm. 
\subsubsection{Comparison against State-of-the-Art.}
We now compare our algorithms against two state-of-the-art algorithms, \texttt{DyTwoSwap} and \texttt{DgOracleTwo}, as well as two greedy strategies. The \texttt{Greedy} strategy is similar to the \texttt{DegGreedy} algorithm described in Section~\ref{subsec:fullydynamic}, with a key difference in how it handles the initial removal of conflicted nodes. Instead of basing this removal on $w(v)/w(N(v))$, it uses the weights of the nodes directly (without $w(N(v))$). Specifically, if an edge is inserted between two independent set nodes, the lighter one is removed from the independent set. In the case of cardinality, which this section addresses, this translates to removing a random node from the two newly adjacent nodes, followed by a simple augmentation step which tries to add adjacent nodes of the node that we just removed from the independent set node. 

\begin{figure}[t]
\centering
\vspace*{-.5cm}
\hspace*{0.3cm}
\includegraphics[width=0.48\textwidth]{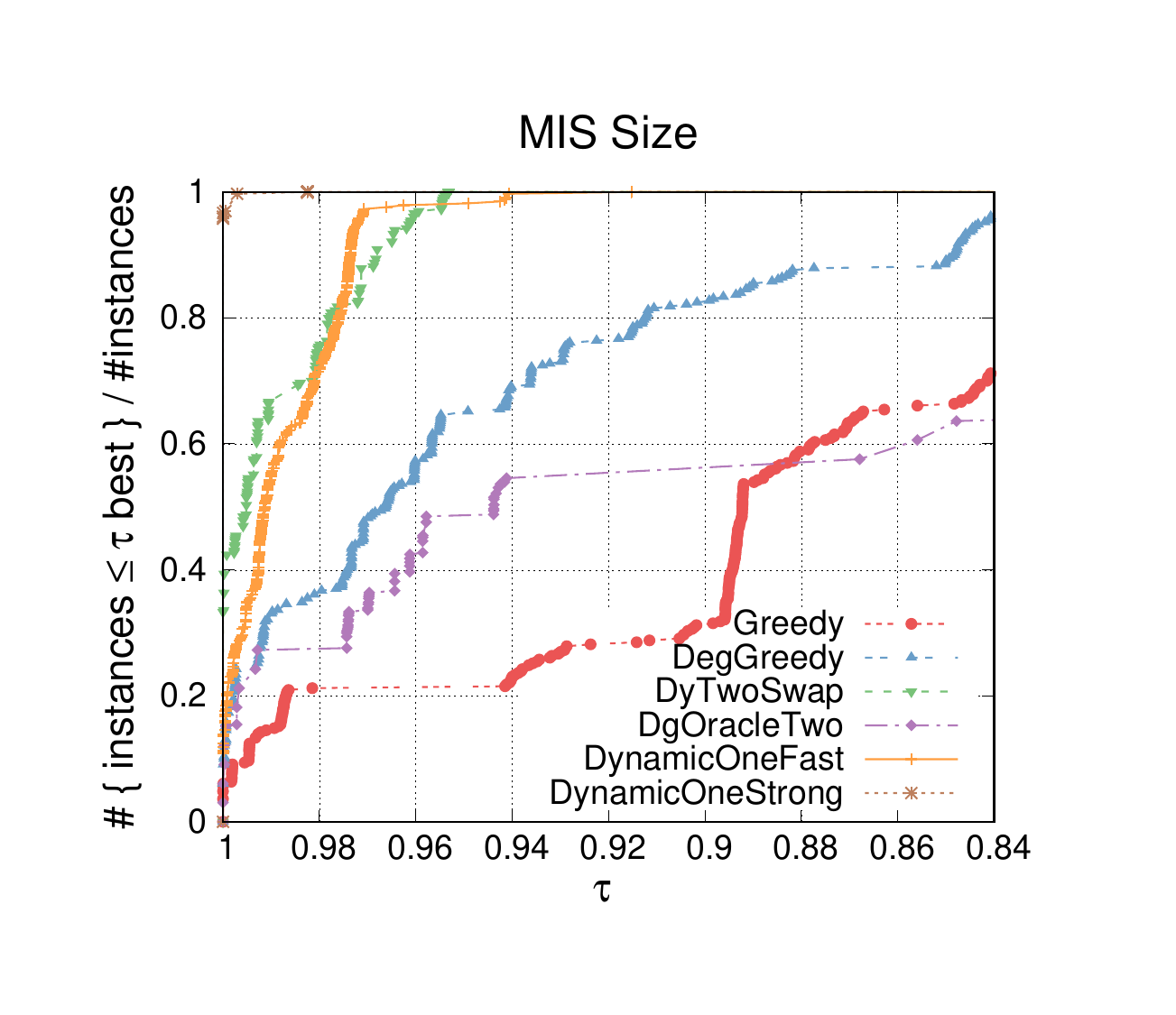} \\
                       \vspace*{-5.5cm}
\includegraphics[width=0.48\textwidth]{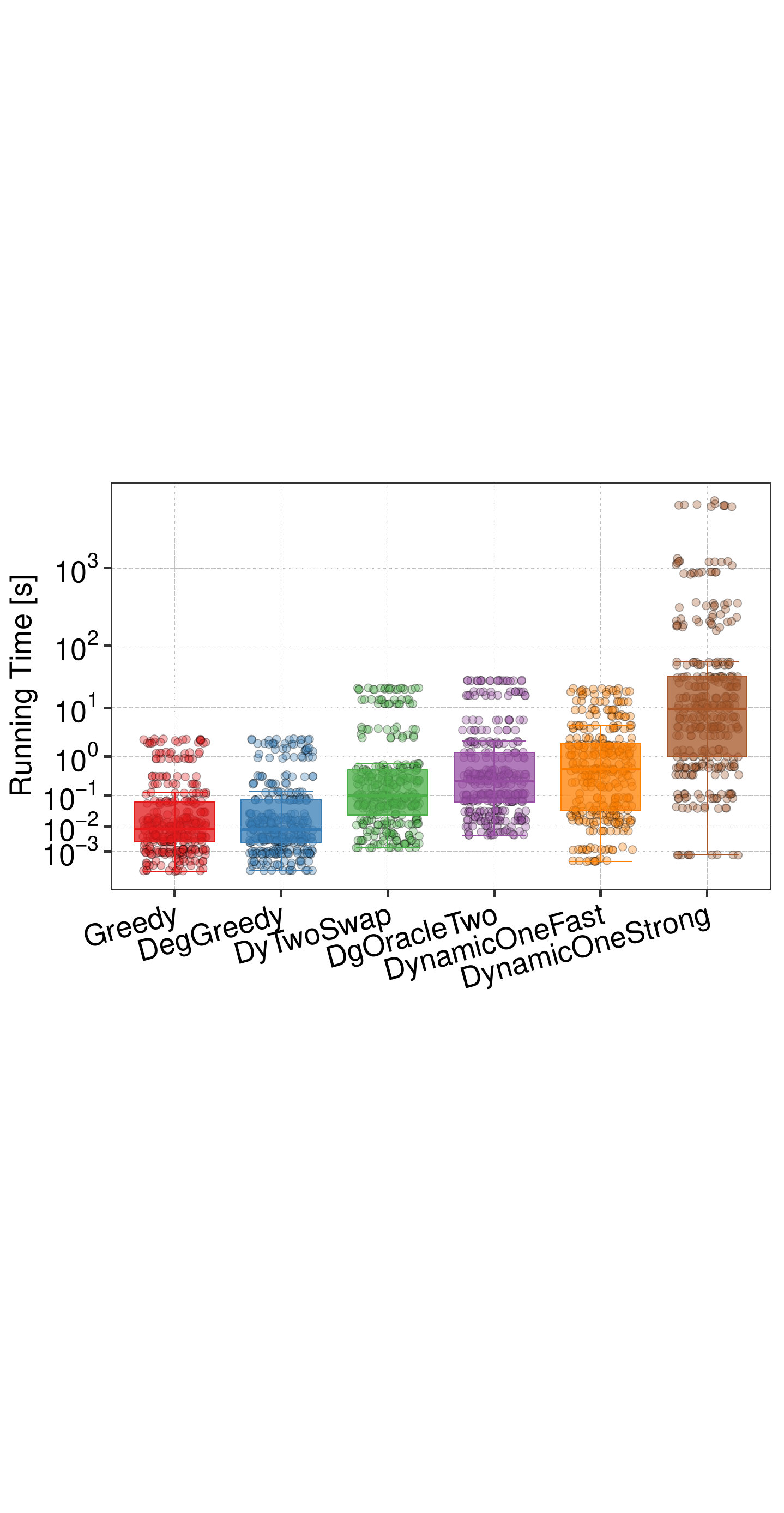}
\vspace*{-5.5cm}
\caption{Performance profile for solution quality (top) and total update time comparison (bottom) against state-of-the-art algorithms on unweighted instances.}
\label{fig:compstateoftheart}
\end{figure}

Figure~\ref{fig:compstateoftheart} gives a performance profile as well as a running time box plot for the various algorithms under consideration. Detailed per instance results can be found in Appendix Table~\ref{table:algorithm_results}. As \texttt{DgOracleTwo} computes worse results than \texttt{DyTwoSwap} and is slower, we only discuss results comparing to \texttt{DyTwoSwap}.

\npthousandsep{\,}
First it can be seen, that our strong algorithm \texttt{DynamicOneStrong} outperforms all other algorithms in terms of solution quality. On 31 out of 33 instances our algorithm computes the best (or equal to best) result. Improvements vary largely on the type of instances that we consider. On mesh type networks, which are generally known to be hard instances for the independent set problem, our method performs particularly well. The largest observed improvement is 3.6\% on the \Id{whitaker3} instance. The good performance is due to the fact that the local subgraphs/problems align closely with the algorithm's intuition, specifically forming a ball around the recently inserted or deleted edge, and the fact that this is a powerful local search operation for the value of $d$ we are using. Similarly, on road networks our algorithm performs well. Our algorithm computes a strictly better result than all of the competitors. The largest observed improvement of \texttt{DynamicOneStrong} over \texttt{DyTwoSwap} is 2.4\% on the \Id{uk} instance.
On social networks, \texttt{DyTwoSwap} and our strong algorithm \texttt{DynamicOneStrong} mostly compute the same results. Initially, this looks a bit surprising, however, the authors already show in their paper that on this type of (easy) instances, their algorithm computes solutions very close to optimal or even optimal solutions. 
More precisely, we also compare algorithms under consideration to the optimum result (see Table~\ref{table:algorithm_results}). We obtained the optimum result after all updates have been performed using the optimum branch-and-reduce solvers of the KaMIS framework. On the instances that the static branch-and-reduce algorithm could solve within a two day time limit, our \texttt{DynamicOneStrong} algorithm \hbox{is at most 0.4\% worse.}

Based on the performance profile, the performance of the \texttt{DynamicOneFast} algorithm in terms of solution quality is comparable to, albeit slightly worse than, the \texttt{DyTwoSwap} algorithm. In terms of running time, it is important to note, however, that \texttt{DynamicOneFast} is slower because it employs a more generic approach for updating the independent sets.
Moreover, our \texttt{DynamicOneStrong} algorithm, although significantly slower than \texttt{DyTwoSwap}, remains orders of magnitude faster than computing independent sets from scratch. To illustrate this, we relate the running time per update of \texttt{DynamicOneStrong} with the time required by the iterated local search algorithm in the KaMIS framework. This comparison is made on the static counterpart of the final instance after all updates. On average (geometric mean), the time per update for \texttt{DynamicOneStrong} is \numprint{1776} times lower than computing the independent set from scratch using weighted iterated local search on the final instance. Note however that this is only a rough estimate as the instances throughout the update \hbox{sequence increase in size.}
\textit{Summing up}, if solution quality is paramount and the structure of the networks is similar to road networks or mesh type networks, \texttt{DynamicOneStrong} is the best choice. If even better quality is required in the respective applications larger values of $d$ and $\nu_\text{max}$ may be feasible. Since our algorithm computes the same results as \texttt{DyTwoSwap} on social networks, but is much slower, \texttt{DyTwoSwap} is still the way to go \hbox{for this type of instances.}

The simple greedy strategies \texttt{Greedy} and \texttt{DegGreedy} are the fastest algorithms, but they are also worse in terms of solution quality. The \texttt{DegGreedy} algorithm performs significantly better than the simpler \texttt{Greedy} algorithm while having a similar running time. On average, the \texttt{DegGreedy} algorithm computes solutions that are 4.2\% smaller (worse) than \texttt{DyTwoSwap}, however, it is also a factor 11.7 faster. Surprisingly, already the simple \texttt{DegGreedy} strategy outperforms the \texttt{DgOracleTwo} algorithm significantly. It computes 70.6\% larger independent sets on average while also being a factor 21 faster. The large improvement mostly stems from the mesh type networks, where the \texttt{DgOracleTwo} algorithm performs particularly bad. In their original paper, the authors did not consider mesh type networks. \textit{Summing up}, the \texttt{DegGreedy} algorithm is a good choice if running time is paramount and solution \hbox{quality is less important.}

\subsection{Fully Dynamic Maximum Weight Independent Set Algorithms.} Another advantage of our algorithm is that it also works with weighted graphs. We are not aware of any other algorithm that can handle general weighted graphs in the fully dynamic setting (\texttt{DyTwoSwap} and \texttt{DgOracleTwo} can not handle weights). Hence, we compare \texttt{DynamicOne} against the simple greedy strategies incorporating weights as well as optimum results.
For this experiment, we run all algorithms on all instances, assigning each vertex a random weight uniformly distributed within the interval [1, 100]. 

\begin{figure}[t]
\centering
\vspace*{-1cm}
\hspace*{0.3cm}
\includegraphics[width=0.48\textwidth]{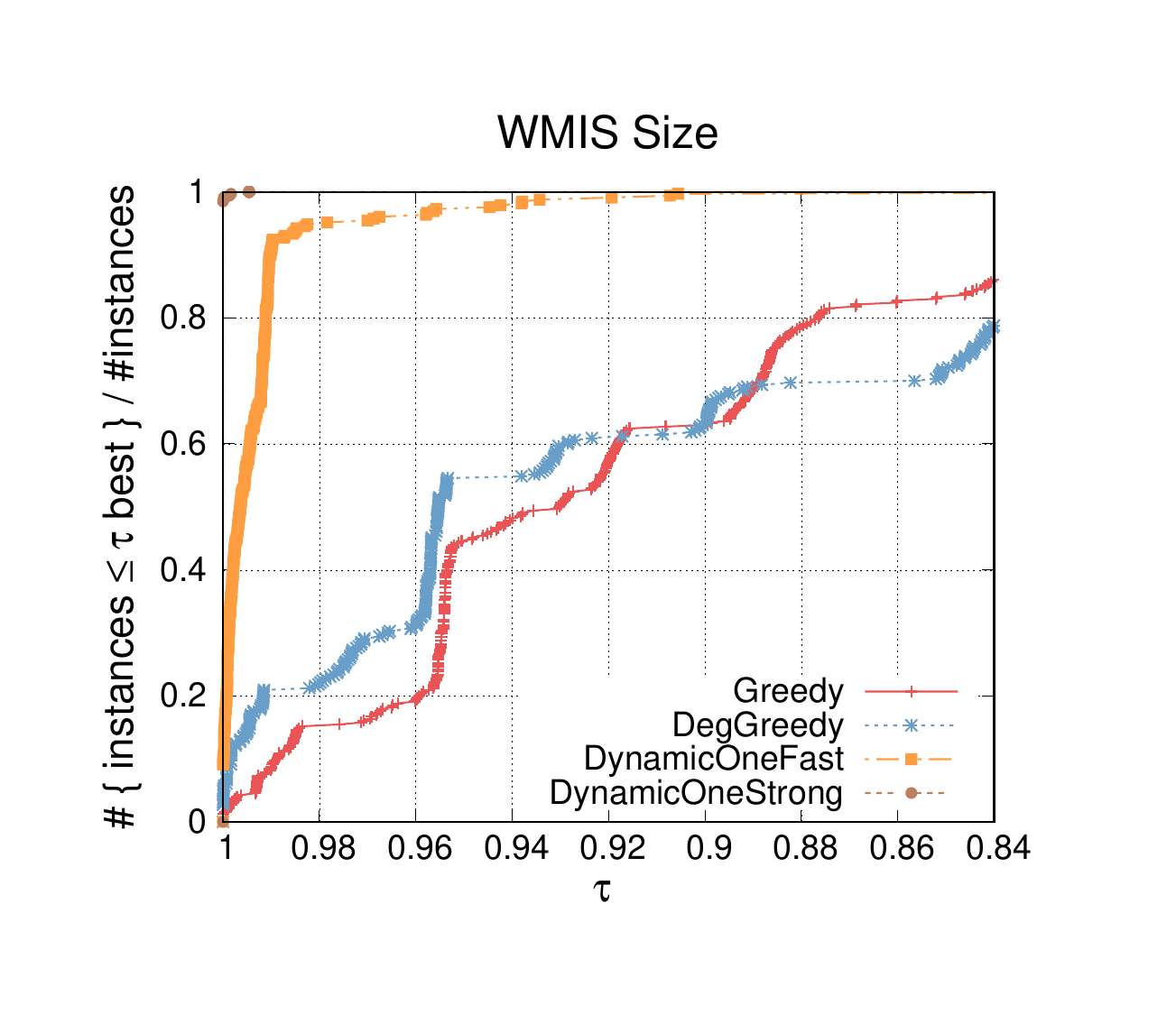} \\
                       \vspace*{-1.5cm}
\includegraphics[width=0.48\textwidth]{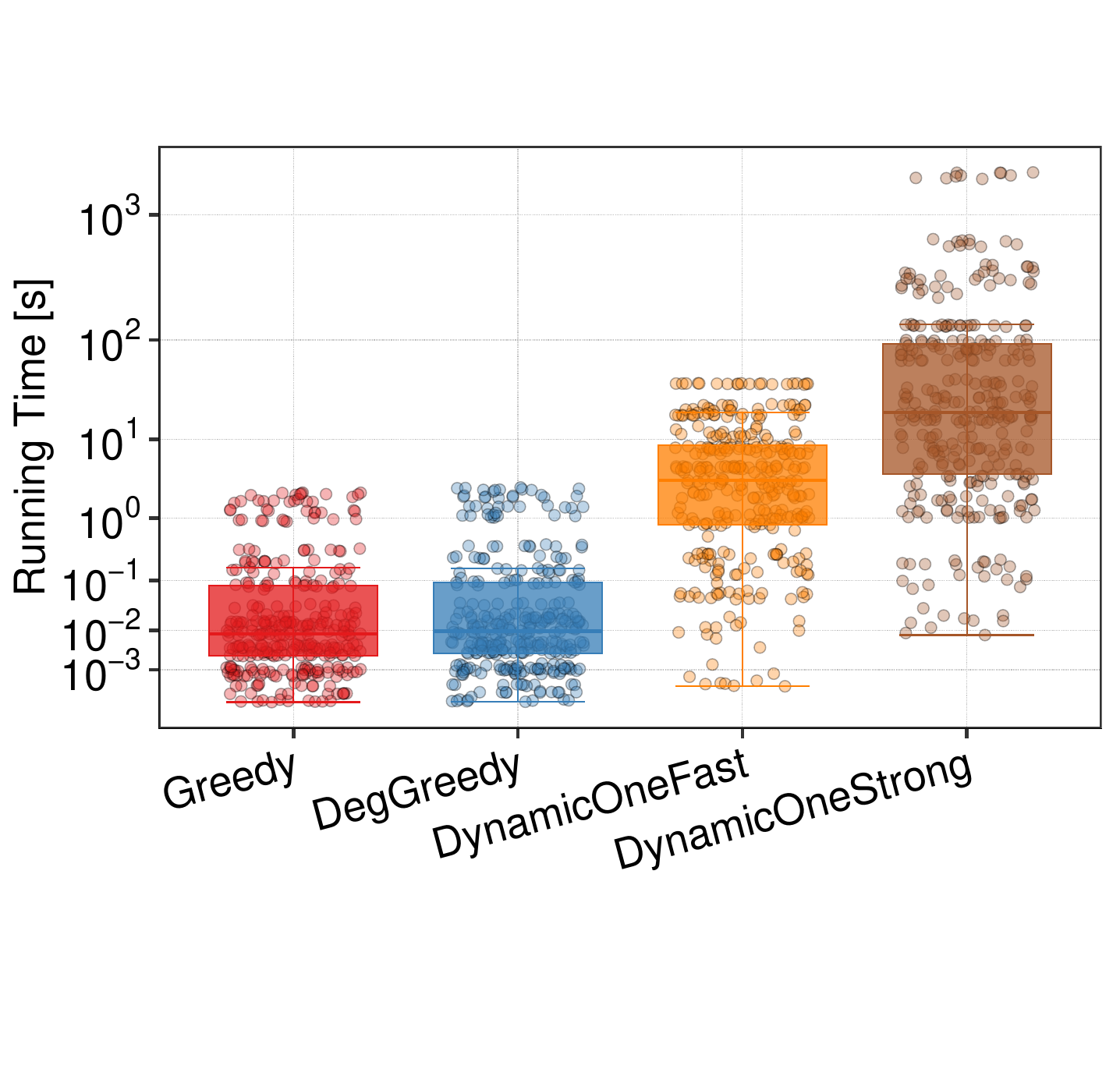}
\vspace*{-1.5cm}
\caption{Performance profile for solution quality (top) and running time box-plot comparison (bottom) against greedy algorithms on weighted instances.}
\label{fig:compstateoftheartweighted}
\end{figure}

Figure~\ref{fig:compstateoftheartweighted} gives a performance profile as well as a running time box plot for the various algorithms under consideration. Detailed per instance results can be found in Appendix Table~\ref{table:algorithm_results_weighted}. Overall, the results are in line with the cardinality case. Our \texttt{DynamicOne} algorithms compute much better solutions than the greedy strategies, but are also slower.  Both of our algorithms outperform \texttt{Greedy} and \texttt{DegGreedy} across all instances. On average, \texttt{DynamicOneFast} and \texttt{DynamicOneStrong} compute solutions that are 8.2\% and 9\% better than \texttt{DegGreedy}, respectively. In contrast to the cardinality case, \texttt{Greedy} and \texttt{DegGreedy} compute mostly similar results. The difference in solution quality \hbox{is less than 0.1\% on average.}

As in the cardinality case, we relate the running time per update of \texttt{DynamicOne} with the time required by the iterated local search algorithm in the KaMIS framework. This comparison is made on the static counterpart of the final instance after all updates. On average (geometric mean), the time per update for \texttt{DynamicOneStrong} and \texttt{DynamicOneFast} is 655 and \numprint{6070} times lower (better), respectively, than computing the independent set from scratch using iterated local search on the final instance. 

Similar to the cardinality case, we compare algorithms under consideration to the optimum result (see Table~\ref{table:algorithm_results_weighted_singleseed}). On the instances that the optimum algorithm from the KaMIS framework could solve to optimality within a two day time limit, our \texttt{DynamicOneStrong} algorithm is at most 0.06\% worse. 

\section{Conclusion}
This paper introduces a novel approach for solving the dynamic maximum weight and cardinality independent set problems using optimal neighborhood exploration. Our method efficiently handles large-scale dynamic graphs typical in real-world applications.
We developed a local search technique that forms and solves independent subproblems to optimality, significantly improving solution quality. Experiments show that our algorithm, featuring a tunable subproblem size parameter, outperforms existing state-of-the-art solvers. Adjusting this parameter allows balancing between running time and solution quality, with increased values leading to better solutions.
In summary, our dynamic optimal neighborhood exploration technique advances the state-of-the-art for dynamic independent set problems, providing a practical and scalable solution for large, dynamic graphs. Future work will focus on further optimization, exploring additional applications in other dynamic graph problems as well as using the technique as a general (parallelizable) local search technique in the static case.

\textbf{Acknowledgments.}
We acknowledge support by DFG grant SCHU 2567/3-1. Moreover, we like to acknowledge Dagstuhl Seminar 22461 on dynamic graph algorithms. %

\vfill \pagebreak
\bibliographystyle{abbrv}
\bibliography{compactfixed}

\clearpage
\appendix

\npthousandsep{\,}
\begin{table*}[h!]
      \centering
      \vspace*{-1cm}

      \begin{tabular}{l@{\hskip 40pt}r@{\hskip 20pt}r}
              \toprule

\multicolumn{3}{c}{Mesh Type Networks} \\
              \midrule
       graph & $n$ & $\mathcal{O}$ \\
                \midrule

\Id{3elt}                              & \numprint{4720}     & \numprint{13722}     \\
\Id{4elt}                              & \numprint{15606}    & \numprint{45878}     \\
\Id{add20}                             & \numprint{2395}     & \numprint{7462}      \\
\Id{add32}                             & \numprint{4960}     & \numprint{9462}      \\
\Id{citeulike_ui}                      & \numprint{731770}   & \numprint{842421}    \\
\Id{crack}                             & \numprint{10240}    & \numprint{30380}     \\
\Id{cs4}                               & \numprint{22499}    & \numprint{43858}     \\
\Id{cti}                               & \numprint{16840}    & \numprint{48232}     \\
\Id{data}                              & \numprint{2851}     & \numprint{15093}     \\
\Id{fe_4elt2}                          & \numprint{11143}    & \numprint{32818}     \\
\Id{fe_body}                           & \numprint{45087}    & \numprint{163734}    \\
\Id{fe_ocean}                          & \numprint{143437}   & \numprint{409593}    \\
\Id{fe_pwt}                            & \numprint{36519}    & \numprint{144794}    \\
\Id{fe_sphere}                         & \numprint{16386}    & \numprint{49152}     \\
\Id{t60k.graph}                        & \numprint{60005}    & \numprint{89440}     \\
\Id{whitaker3.graph}                   & \numprint{9800}     & \numprint{28989}     \\
\Id{wing.graph}                        & \numprint{62032}    & \numprint{121544}    \\

\midrule
\multicolumn{3}{c}{Road Networks} \\
\midrule
graph                                  & $n$                 & $\mathcal{O}$        \\
\midrule
\Id{asia.osm}                          & \numprint{11950757} & \numprint{12711603}  \\
\Id{belgium.osm}                       & \numprint{1441295}  & \numprint{1549970}   \\
\Id{germany.osm}                       & \numprint{11548845} & \numprint{12369181}  \\
\Id{great-britain.osm}                 & \numprint{7733822}  & \numprint{8156517}   \\
\Id{italy.osm.graph}                   & \numprint{6686493}  & \numprint{7013978}   \\
\Id{luxembourg.osm.graph}              & \numprint{114599}   & \numprint{119666}    \\
\Id{netherlands.osm.graph}             & \numprint{2216688}  & \numprint{2441238}   \\
\Id{uk.graph}                          & \numprint{4824}     & \numprint{6837}      \\
\midrule
\multicolumn{3}{c}{Social Networks} \\
\midrule
graph                                  & $n$                 & $\mathcal{O}$        \\
\midrule
\Id{amazon-ratings}                    & \numprint{2146058}  & \numprint{477676}     \\
\Id{dnc-temporalGraph}                 & \numprint{2030}     & \numprint{4384}      \\
\Id{facebook-wosn-wall}                & \numprint{46953}    & \numprint{183412}    \\
\Id{haggle}                            & \numprint{275}      & \numprint{2124}      \\
\Id{lastfm_band}                       & \numprint{174078}   & \numprint{894388}    \\
\Id{lkml-reply}                        & \numprint{63400}    & \numprint{159996}    \\
\Id{sociopatterns-infections}          & \numprint{411}      & \numprint{2765}      \\
\Id{topology}                          & \numprint{34762}    & \numprint{107720}    \\

      \bottomrule
      \end{tabular}
      \caption{Basic properties of benchmark set of static and real dynamic graphs from~\cite{benchmarksfornetworksanalysis,DBLP:journals/corr/abs-2003-00736,UFsparsematrixcollection,snap,DBLP:conf/www/Kunegis13,konect:unlink,DBLP:journals/jpdc/FunkeLMPSSSL19,kappa}.
We report the original number update operations $\mathcal{O}$, after removing obsolete updates (such as parallel edges, self-loops etc.). Note that most of these instances only feature insertions.  } 

      \label{staticgraphs}
\label{dyninstances}
\label{tab:graphstable}
\end{table*}

\begin{landscape}

\begin{table}[h!]
\centering
\vspace*{-1cm}
\begin{tabular}{l@{\hskip 20pt}r@{\hskip 20pt}r@{\hskip 20pt}r@{\hskip 20pt}r@{\hskip 20pt}||r@{\hskip 20pt}r@{\hskip 20pt}||r@{\hskip 20pt}}
\toprule
\texttt{Graph} &  \texttt{Greedy} & \texttt{DegGreedy} & \texttt{DyTwoSwap} & \texttt{DgOracleTwo} & \texttt{DynOneFast} & \texttt{DynOneStrong} & \texttt{OPT}\\
\midrule \multicolumn{8}{c}{Mesh Networks} \\ 
\midrule

\Id{3elt}      & \numprint{1240}  & \numprint{1349}          & \numprint{1442}          & \numprint{78}    & \numprint{1471}           & \textbf{\numprint{1475}} & $\times$\\
\Id{4elt}      & \numprint{4048}  & \numprint{4389}          & \numprint{4777}          & \numprint{420}   & \numprint{4905}           & \textbf{\numprint{4918}} & $\times$\\
\Id{add20}     & \numprint{1033}  & \numprint{1100}          & \textbf{\numprint{1130}} & \numprint{958}   & \numprint{1121}           & \textbf{\numprint{1130}} & \numprint{1130}\\
\Id{add32}     & \numprint{2141}  & \numprint{2269}          & \textbf{\numprint{2286}} & \numprint{1984}  & \numprint{2269}           & \textbf{\numprint{2286}} & \numprint{2286} \\
\Id{crack}     & \numprint{4128}  & \numprint{4572}          & \numprint{4602}          & \numprint{4572}  & \numprint{4572}           & \textbf{\numprint{4603}} & \numprint{4603}\\
\Id{cs4}       & \numprint{7708}  & \numprint{7769}          & \numprint{8767}          & \numprint{7536}  & \numprint{8964}           & \textbf{\numprint{9138}} & $\times$\\
\Id{cti}       & \numprint{6155}  & \numprint{7552}          & \numprint{7856}          & \numprint{1998}  & \textbf{\numprint{8088}}  & \textbf{\numprint{8088}} & $\times$\\
\Id{data}      & \numprint{637}   & \numprint{678}           & \textbf{\numprint{685}}  & \numprint{105}   & \numprint{678}            & \numprint{683} & $\times$\\
\Id{fe_4elt2}  & \numprint{2904}  & \numprint{3008}          & \numprint{3448}          & \numprint{1166}  & \numprint{3505}           & \textbf{\numprint{3562}} & $\times$\\
\Id{fe_body}   & \numprint{12012} & \numprint{12760}         & \numprint{13465}         & \numprint{4207}  & \numprint{13620}          & \textbf{\numprint{13730}} & $\times$\\
\Id{fe_ocean}  & \numprint{53444} & \numprint{70456}         & \numprint{70514}         & \numprint{6914}  & \textbf{\numprint{71664}} & \numprint{71629} & \numprint{71716}\\
\Id{fe_pwt}    & \numprint{7806}  & \numprint{8683}          & \numprint{9033}          & \numprint{3135}  & \numprint{9038}           & \textbf{\numprint{9219}} & $\times$\\
\Id{fe_sphere} & \numprint{4230}  & \textbf{\numprint{5462}} & \textbf{\numprint{5462}} & \numprint{655}   & \textbf{\numprint{5462}}  & \textbf{\numprint{5462}} & $\times$\\
\Id{t60k}      & \numprint{24158} & \numprint{29403}         & \numprint{29621}         & \numprint{29432} & \numprint{29437}          & \textbf{\numprint{29644}} & $\times$\\
\Id{whitaker3} & \numprint{2515}  & \numprint{2646}          & \numprint{3002}          & \numprint{612}   & \numprint{3095}           & \textbf{\numprint{3113}} & $\times$\\
\Id{wing}      & \numprint{21089} & \numprint{20784}         & \numprint{23967}         & \numprint{20868} & \numprint{24612}          & \textbf{\numprint{25144}} & $\times$ \\
\midrule \multicolumn{8}{c}{Road Networks} \\ 
\midrule

\Id{asia.osm}          & \numprint{5337245} & \numprint{5806366} & \numprint{5951596} & \numprint{5767404} & \numprint{5874844} & \textbf{\numprint{5980482}} & \numprint{5998335}\\
\Id{belgium.osm}       & \numprint{645260}  & \numprint{689483}  & \numprint{716522}  & \numprint{693765}  & \numprint{703848}  & \textbf{\numprint{721749}} & \numprint{724487}\\
\Id{germany.osm}       & \numprint{5210948} & \numprint{5569050} & \numprint{5780252} & \numprint{5575465} & \numprint{5687811} & \textbf{\numprint{5820972}} & \numprint{5841600}\\
\Id{great-britain.osm} & \numprint{3535628} & \numprint{3789940} & \numprint{3921417} & \numprint{3782883} & \numprint{3849252} & \textbf{\numprint{3946495}} & \numprint{3958619}\\
\Id{italy.osm}         & \numprint{2983420} & \numprint{3253962} & \numprint{3328962} & \numprint{3257039} & \numprint{3265390} & \textbf{\numprint{3342945}} & \numprint{3353582}\\
\Id{luxembourg.osm}    & \numprint{51428}   & \numprint{55476}   & \numprint{57123}   & \numprint{55923}   & \numprint{56167}   & \textbf{\numprint{57424}} & \numprint{57663}\\
\Id{netherlands.osm}   & \numprint{994501}  & \numprint{1042427} & \numprint{1102700} & \numprint{1050646} & \numprint{1084286} & \textbf{\numprint{1113242}} & \numprint{1116770}\\
\Id{uk}                & \numprint{1937}    & \numprint{2006}    & \numprint{2140}    & \numprint{2063}    & \numprint{2170}    & \textbf{\numprint{2192}} & \numprint{2198}\\
\midrule 
\multicolumn{8}{c}{Social Networks} \\
\midrule
\Id{amazon-ratings}            & \numprint{1982026}         & \numprint{1991436}      & \textbf{\numprint{1992583}} & \numprint{1992071}      & \numprint{1991669}         & \numprint{1992582} & \numprint{1992590}\\
\Id{citeulike_ui}              & \numprint{708835}          & \numprint{709906}       & \textbf{\numprint{710065}}  & \numprint{709841}       & \numprint{709985}          & \textbf{\numprint{710065}} & \numprint{710067}\\
\Id{dnc-temporalGraph}         & \numprint{1773}            & \numprint{1780}         & \textbf{\numprint{1781}}    & \numprint{1775}         & \numprint{1780}            & \textbf{\numprint{1781}} & \numprint{1781}\\
\Id{facebook-wosn-wall}        & \numprint{23651}           & \numprint{24498}        & \numprint{25054}            & \numprint{24358}        & \numprint{24945}           & \textbf{\numprint{25121}} & \numprint{25126}\\
\Id{haggle}                    & \textbf{\numprint{234}}    & \textbf{\numprint{234}} & \textbf{\numprint{234}}     & \textbf{\numprint{234}} & \textbf{\numprint{234}}    & \textbf{\numprint{234}} & \numprint{234}\\
\Id{lastfm_band}               & \textbf{\numprint{173087}} & \numprint{173066}       & \textbf{\numprint{173087}}  & \numprint{173086}       & \textbf{\numprint{173087}} & \textbf{\numprint{173087}} & \numprint{173087}\\
\Id{lkml-reply}                & \numprint{53418}           & \numprint{53926}        & \numprint{54053}            & \numprint{53909}        & \numprint{54024}           & \textbf{\numprint{54063}} & \numprint{54065}\\
\Id{sociopatterns-infections}  & \numprint{110}             & \numprint{114}          & \textbf{\numprint{118}}     & \numprint{101}          & \textbf{\numprint{118}}    & \textbf{\numprint{118}} & \numprint{118}\\
\Id{topology}                  & \numprint{29224}           & \numprint{29504}        & \textbf{\numprint{29565}}   & \numprint{29548}        & \numprint{29504}           & \textbf{\numprint{29565}} & \numprint{29565}\\

\bottomrule
\end{tabular}
\caption{Detailed per instance results (\textit{higher} is better). Each entry corresponds to the best result found out of ten repetitions for each algorithm after all updates have been performed. The best result is highlighted in \textbf{bold} font. Here, we also report the optimum value, if the optimum solver found the optimum solution within a 48h time limit. If the solver did not find an optimum solution in that time, we put in a cross $\times$ in the respective field.}
\label{table:algorithm_results}
\end{table}

\end{landscape}

\begin{landscape}
\nprounddigits{0}
\begin{table}[h!]
\centering
\vspace*{-0.5cm}
\begin{tabular}{l@{\hskip 20pt}r@{\hskip 20pt}r@{\hskip 20pt}||r@{\hskip 20pt}r@{\hskip 20pt}r@{\hskip 20pt}r}
\toprule
\texttt{Graph} &  \texttt{Greedy} & \texttt{DegGreedy} & \texttt{DynOneFast} & \texttt{DynOneStrong} \\
\midrule \multicolumn{5}{c}{Mesh Networks} \\ 
\midrule
\Id{3elt} & \numprint{80427.10} & \numprint{75634.70} & \numprint{90111.70} & \textbf{\numprint{90169.10}} \\ 
\Id{4elt} & \numprint{263808.50} & \numprint{245425.90} & \numprint{296243.80} & \textbf{\numprint{296483.60}} \\ 
\Id{add20} & \numprint{66379.30} & \numprint{67344.40} & \numprint{69128.30} & \textbf{\numprint{69357.30}} \\ 
\Id{add32} & \numprint{133652.60} & \numprint{138037.40} & \numprint{140734.70} & \textbf{\numprint{141074.20}} \\ 
\Id{crack} & \numprint{201774.00} & \numprint{235130.80} & \numprint{240666.00} & \textbf{\numprint{241509.50}} \\ 
\Id{cs4} & \numprint{505079.30} & \numprint{493488.60} & \numprint{545842.20} & \textbf{\numprint{549019.90}} \\ 
\Id{cti} & \numprint{339123.30} & \numprint{350490.70} & \numprint{414558.10} & \textbf{\numprint{416152.50}} \\ 
\Id{data} & \numprint{34846.50} & \numprint{35189.40} & \numprint{41695.90} & \textbf{\numprint{43953.90}} \\ 
\Id{fe_4elt2} & \numprint{191496.20} & \numprint{179535.70} & \numprint{211390.40} & \textbf{\numprint{214538.50}} \\ 
\Id{fe_body} & \numprint{734140.50} & \numprint{703939.80} & \numprint{834556.60} & \textbf{\numprint{838087.10}} \\ 
\Id{fe_ocean} & \numprint{2910035.70} & \numprint{3054326.20} & \numprint{3584669.80} & \textbf{\numprint{3610587.90}} \\ 
\Id{fe_pwt} & \numprint{515032.30} & \numprint{472183.70} & \numprint{584843.30} & \textbf{\numprint{620905.20}} \\ 
\Id{fe_sphere} & \numprint{272750.70} & \numprint{257150.00} & \numprint{308532.00} & \textbf{\numprint{308626.70}} \\ 
\Id{t60k} & \numprint{1498465.30} & \numprint{1514035.50} & \numprint{1623871.00} & \textbf{\numprint{1626908.50}} \\ 
\Id{whitaker3} & \numprint{164853.00} & \numprint{152056.90} & \numprint{185312.90} & \textbf{\numprint{185694.20}} \\ 
\Id{wing} & \numprint{1388751.20} & \numprint{1360853.10} & \numprint{1499895.00} & \textbf{\numprint{1512689.00}} \\ 

\midrule \multicolumn{5}{c}{Road Networks} \\ 
\midrule

\Id{asia.osm} & \numprint{337151643.10} & \numprint{337879210.40} & \numprint{350337475.60} & \textbf{\numprint{353127303.00}} \\ 
\Id{belgium.osm} & \numprint{40625661.40} & \numprint{40687448.60} & \numprint{42226434.60} & \textbf{\numprint{42590970.80}} \\ 
\Id{germany.osm} & \numprint{327151985.70} & \numprint{327763791.40} & \numprint{339946810.60} & \textbf{\numprint{342976041.50}} \\ 
\Id{great-britain.osm} & \numprint{220872346.10} & \numprint{221522336.40} & \numprint{229334871.70} & \textbf{\numprint{231482436.80}} \\ 
\Id{italy.osm} & \numprint{188795345.70} & \numprint{189286892.70} & \numprint{196001471.40} & \textbf{\numprint{197612298.80}} \\ 
\Id{luxembourg.osm} & \numprint{3241919.10} & \numprint{3248699.90} & \numprint{3359823.20} & \textbf{\numprint{3393258.30}} \\ 
\Id{netherlands.osm} & \numprint{62461275.20} & \numprint{62507042.00} & \numprint{64922456.30} & \textbf{\numprint{65544508.50}} \\ 
\Id{uk} & \numprint{122900.00} & \numprint{122093.30} & \numprint{130282.40} & \textbf{\numprint{130875.20}} \\

\midrule 
\multicolumn{5}{c}{Social Networks} \\
\midrule

\Id{amazon-ratings} & \numprint{101520569.20} & \numprint{102066076.30} & \numprint{102124692.50} & \textbf{\numprint{102237718.00}} \\ 
\Id{citeulike_ui} & \numprint{35600375.40} & \numprint{35951479.30} & \numprint{35970737.10} & \textbf{\numprint{35979609.30}} \\ 
\Id{dnc-temporalGraph} & \numprint{89498.10} & \numprint{91018.50} & \numprint{91369.40} & \textbf{\numprint{91427.70}} \\ 
\Id{facebook-wosn-wall} & \numprint{1290378.80} & \numprint{1333313.00} & \numprint{1381790.50} & \textbf{\numprint{1389705.10}} \\ 
\Id{haggle} & \numprint{10979.60} & \numprint{11800.20} & \numprint{11800.20} & \textbf{\numprint{11839.40}} \\ 
\Id{lastfm_band} & \numprint{8714534.30} & \numprint{8730187.50} & \numprint{8731008.30} & \textbf{\numprint{8731010.20}} \\ 
\Id{lkml-reply} & \numprint{2758582.70} & \numprint{2776324.80} & \numprint{2796715.90} & \textbf{\numprint{2799473.30}} \\ 
\Id{sociopatterns-infections} & \numprint{6018.30} & \numprint{6276.70} & \numprint{7015.40} & \textbf{\numprint{7031.50}} \\ 
\Id{topology} & \numprint{1473077.40} & \numprint{1516556.40} & \numprint{1523308.40} & \textbf{\numprint{1525097.20}} \\ 

\bottomrule
\end{tabular}
\caption{Detailed per instance results (\textit{higher} is better) on weighted instances. Each entry corresponds to the average result for each algorithm after all updates have been performed. The best result is highlighted in \textbf{bold} font.}
\label{table:algorithm_results_weighted}
\end{table}

\end{landscape}

\begin{landscape}
\nprounddigits{0}
\begin{table}[h!]
\centering
\vspace*{-0.5cm}
\begin{tabular}{l@{\hskip 20pt}r@{\hskip 20pt}r@{\hskip 20pt}||r@{\hskip 20pt}r||@{\hskip 20pt}r@{\hskip 20pt}r}
\toprule
\texttt{Graph} &  \texttt{Greedy} & \texttt{DegGreedy} & \texttt{DynOneFast} & \texttt{DynOneStrong} & OPT\\
\midrule \multicolumn{5}{c}{Mesh Networks} \\ 
\midrule

\Id{3elt}                                     & \numprint{80433.00}     & \numprint{76471.00}          & \numprint{90825.00}            & \textbf{\numprint{90909.00}}     & \numprint{90924}\\
\Id{4elt}                                     & \numprint{262673.00}    & \numprint{244985.00}         & \numprint{295890.00}           & \textbf{\numprint{296240.00}}    & \numprint{296316} \\
\Id{add20}                                    & \numprint{66623.00}     & \numprint{67807.00}          & \numprint{69524.00}            & \textbf{\numprint{69672.00}}     & \numprint{69672}\\
\Id{add32}                                    & \numprint{132463.00}    & \numprint{138179.00}         & \numprint{140864.00}           & \textbf{\numprint{141182.00}}    & \numprint{141259}\\
\Id{crack}                                    & \numprint{204346.00}    & \numprint{236599.00}         & \numprint{241800.00}           & \textbf{\numprint{242904.00}}    & \numprint{242922}\\
\Id{cs4}                                      & \numprint{503367.00}    & \numprint{494685.00}         & \numprint{546236.00}           & \textbf{\numprint{549342.00}}    & $\times$\\
\Id{cti}                                      & \numprint{339217.00}    & \numprint{349883.00}         & \numprint{416277.00}           & \textbf{\numprint{420197.00}}    & $\times$\\
\Id{data}                                     & \numprint{34914.00}     & \numprint{34035.00}          & \numprint{41839.00}            & \textbf{\numprint{44602.00}}     & $\times$\\
\Id{fe_4elt2}                                 & \numprint{192996.00}    & \numprint{181627.00}         & \numprint{211970.00}           & \textbf{\numprint{215739.00}}    & \numprint{215755}\\
\Id{fe_body}                                  & \numprint{730297.00}    & \numprint{708021.00}         & \numprint{836208.00}           & \textbf{\numprint{840744.00}}    & $\times$\\
\Id{fe_ocean}                                 & \numprint{2908069.00}   & \numprint{3034363.00}        & \numprint{3575131.00}          & \textbf{\numprint{3601335.00}}   & $\times$\\
\Id{fe_pwt}                                   & \numprint{516239.00}    & \numprint{472353.00}         & \numprint{584247.00}           & \textbf{\numprint{645158.00}}    & $\times$\\
\Id{fe_sphere}                                & \numprint{274290.00}    & \numprint{258583.00}         & \numprint{309310.00}           & \textbf{\numprint{309338.00}} &  $\times$\\
\Id{t60k}                                     & \numprint{1494694.00}   & \numprint{1508797.00}        & \numprint{1622343.00}          & \textbf{\numprint{1624919.00}}   & \numprint{1625636}\\
\Id{whitaker3}                                & \numprint{166549.00}    & \numprint{152877.00}         & \numprint{186574.00}           & \textbf{\numprint{187048.00}}    & \numprint{187150}\\
\Id{wing}                                     & \numprint{1389365.00}   & \numprint{1360105.00}        & \numprint{1500520.00}          & \textbf{\numprint{1513556.00}}   & $\times$\\

\midrule \multicolumn{5}{c}{Road Networks} \\
\midrule

\Id{asia.osm}                                 & \numprint{337172858.00} & \numprint{337881640.00}      & \numprint{350340186.00}        & \textbf{\numprint{353137274.00}} & \numprint{353174593}\\
\Id{belgium.osm}                              & \numprint{40641709.00}  & \numprint{40701734.00}       & \numprint{42240786.00}         & \textbf{\numprint{42605423.00}}  & \numprint{42610605}\\
\Id{germany.osm}                              & \numprint{327166065.00} & \numprint{327791692.00}      & \numprint{339953614.00}        & \textbf{\numprint{342990960.00}} & \numprint{343041226}\\
\Id{great-britain.osm}                        & \numprint{220893807.00} & \numprint{221517185.00}      & \numprint{229324772.00}        & \textbf{\numprint{231483948.00}} & \numprint{231529608}\\
\Id{italy.osm}                                & \numprint{188808339.00} & \numprint{189295075.00}      & \numprint{196007403.00}        & \textbf{\numprint{197630568.00}} & \numprint{197651869}\\
\Id{luxembourg.osm}                           & \numprint{3229282.00}   & \numprint{3235616.00}        & \numprint{3351233.00}          & \textbf{\numprint{3384719.00}}   & \numprint{3384957}\\
\Id{netherlands.osm}                          & \numprint{62504555.00}  & \numprint{62538118.00}       & \numprint{64944070.00}         & \textbf{\numprint{65568780.00}}  & \numprint{65578800}\\
\Id{uk}                                       & \numprint{122753.00}    & \numprint{122449.00}         & \numprint{130285.00}           & \textbf{\numprint{130899.00}}    & \numprint{130946}\\

\midrule
\multicolumn{5}{c}{Social Networks} \\
\midrule

\Id{amazon-ratings}                           & \numprint{101602864.00} & \numprint{102115654.00}      & \numprint{102186265.00}        & \textbf{\numprint{102287992.00}} & \numprint{102293304} \\
\Id{citeulike_ui}                             & \numprint{35631253.00}  & \numprint{35964477.00}       & \numprint{35983929.00}         & \textbf{\numprint{35994003.00}}  & \numprint{35994022}\\
\Id{dnc-temporalGraph}                        & \numprint{88238.00}     & \numprint{90040.00}          & \numprint{90388.00}            & \textbf{\numprint{90424.00}}     & \numprint{90424}\\
\Id{facebook-wosn-wall}                       & \numprint{1289402.00}   & \numprint{1330707.00}        & \numprint{1378783.00}          & \textbf{\numprint{1388061.00}}   & $\times$\\
\Id{haggle}                                   & \numprint{9852.00}      & \textbf{\numprint{11647.00}} & \textbf{\numprint{11647.00}}   & \textbf{\numprint{11647.00}}     & \numprint{11647}\\
\Id{lastfm_band}                              & \numprint{8698740.00}   & \numprint{8717211.00}        & \textbf{\numprint{8719546.00}} & \textbf{\numprint{8719546.00}}   & \numprint{8719546}\\
\Id{lkml-reply}                               & \numprint{2757010.00}   & \numprint{2775771.00}        & \numprint{2796309.00}          & \textbf{\numprint{2799487.00}}   & \numprint{2799682}\\
\Id{sociopatterns-infections}                 & \numprint{5926.00}      & \numprint{6410.00}           & \numprint{7022.00}             & \textbf{\numprint{7125.00}}      & \numprint{7125}\\
\Id{topology}                                 & \numprint{1463982.00}   & \numprint{1519086.00}        & \numprint{1525776.00}          & \textbf{\numprint{1527491.00}}   & \numprint{1527512}\\

\bottomrule
\end{tabular}
\caption{Detailed per instance results (\textit{higher} is better) on weighted instances. Each entry corresponds to the first seed value (on corresponding weighted instance) result for each algorithm after all updates have been performed. The best result is highlighted in \textbf{bold} font. Here, we also report the optimum value, if the optimum solver found the optimum solution within a 48h time limit. If the solver did not find an optimum solution in that time, we put in a cross $\times$ in the respective field.}
\label{table:algorithm_results_weighted_singleseed}
\end{table}

\end{landscape}

\end{document}

%% file: shortings.tex
\newcommand{\term}[1]{\emph{#1}}
\newcommand{\algoname}[1]{\textsc{#1}}

\newcommand{\baR}[0]{branch-and-reduce\xspace}

\newcommand{\ls}[0]{local-search\xspace}

\newcommand{\ils}[0]{iterated \ls}

\newcommand{\optNeighExplo}[0]{optimal neighborhood exploration\xspace}

\newcommand{\ie}{i.e.\ }
\newcommand{\etal}{et~al.}
\newcommand{\eg}{e.g.\ }

\newcommand{\I}{\mathcal{I}}

\newcommand{\hils}{\textsf{HILS}}
\newcommand{\htwis}{\textsf{HtWIS}}

%% file: math_definitions.tex
\DeclareMathOperator{\wMap}{\ensuremath{\omega}}

\newcommand{\IS}[0]{\ensuremath{\mathcal{I}}}

%% file: makros.tex
\usepackage{amsfonts}

\newcommand{\natnull}{\mathbb{N}_{0}}

\newcommand{\realrange}[2]{\left[#1, #2\right]}

\newcommand{\unitrange}[2]{\realrange{0}{1}}

\newcommand{\llabel}[1]{\label{\labelprefix:#1}}
\newcommand{\labelprefix}{} %

\newcommand{\discussionsize}{\small}

\newenvironment{code}{\noindent%
\begin{tabbing}%
\hspace{2em}\=\hspace{2em}\=\hspace{2em}\=\hspace{2em}\=\hspace{2em}\=%
\hspace{2em}\=\hspace{2em}\=\hspace{2em}\=\hspace{2em}\=\hspace{2em}\=%
\kill}{\end{tabbing}}

\newcommand{\labelcommand}{}
\newcommand{\captiontext}{}
\newsavebox{\codeparam}
\newcounter{lineNumber}
\newenvironment{disscodepos}[3]{%
\renewcommand{\labelcommand}{#2}%
\renewcommand{\captiontext}{#3}%
\sbox{\codeparam}{\parbox{\textwidth}{#3}}%
\begin{figure}[#1]\begin{center}\begin{code}\setcounter{lineNumber}{1}}{%
\end{code}\end{center}\caption{\llabel{\labelcommand}\captiontext}\end{figure}}

{\end{disscodepos}}

\newdimen\endofsize\endofsize=0.5em
\def\endofbeweis{~\quad\hglue\hsize minus\hsize
                 \hbox{\vrule height \endofsize width
\endofsize}\par}

%% file: figures/non_maximal.tex
	\begin{tikzpicture}
		\node[draw, circle, color=green, fill=green, inner sep=2pt, label=right:$3$] (A) at (1,0) {};
		\node[draw, circle, fill=black, inner sep=2pt, label=below:$u$, label=left:$1$] (u) at (1.9,1) {};
		\node[draw, circle, color=green, fill=green, inner sep=2pt, label=below:$3$] (C) at (3.5,1) {};
		\node[draw, circle, fill=orange, color=orange, inner sep=2pt, label=right:$2$, label=above:$v$] (D) at (1,2) {};
		\node[draw, circle, fill=orange, color=orange, inner sep=2pt, label=left:$2$] (E) at (2.75,2) {};
		\node[draw, circle, fill=black, inner sep=2pt] (F) at (-0.5,2) {};
		\node[draw, circle, fill=black, inner sep=2pt] (G) at (4.5,2) {};
		\node[draw, circle, fill=black, inner sep=2pt] (J) at (-.5,1) {};

		\draw (A) -- (u);
		\draw (u) -- (C);
		\draw (A) -- (D);
		\draw (u) -- (E);
		\draw (C) -- (E);
		\draw (D) -- (F);
		\draw (E) -- (G);
		\draw (A) -- (J);
		\draw (J) -- (D);

		\draw[red, thick, dashed] (0.4,-0.5) -- (4,-0.5) -- (4,2.55) -- (0.4,2.55) -- cycle;
		
		\draw[blue, thick, dashed] (0.5,-0.4) -- (3.9,-0.4) -- (3.9,2.45) -- (2.2,2.45) -- (0.5,0.75) -- cycle;
		\node[red] at (2.15,2.8) {$H$};
		
	\end{tikzpicture}